\documentclass[a4paper]{amsart}

\usepackage[utf8]{inputenc}
\usepackage{mathbbol,mathrsfs,amsfonts,amssymb,amsthm,stmaryrd}
\usepackage{amsmath}
\usepackage[all]{xypic}
\usepackage{mathpazo}
\usepackage[mathcal]{euscript}
\usepackage{graphicx, floatflt}
\usepackage{multicol}
\usepackage{sidecap}
\usepackage{nicefrac}
\usepackage{booktabs}
\usepackage{caption,morefloats}

\usepackage{verbatim} 

\usepackage[colorlinks,hyperindex]{hyperref}
\hypersetup{linkcolor=blue, citecolor=blue}

\newcommand{\ins}{{\text{int}}}
\newcommand{\m}{{\,\textrm{m} }}

\newcommand*{\red}[1]{{#1}}
\newcommand*{\orange}[1]{{#1}}
\newcommand*{\gold}[1]{{#1}}

\theoremstyle{theorem}

\begin{document}

\title[The Late Jolt]{The Late Jolt\\ \footnotesize{Re-Examining the World Trade Center Catastrophe}}

\author[Ansgar Schneider]{Ansgar Schneider}

\begin{abstract}

The Twin Towers of the World Trade Center collapsed  in a progressive top to bottom
manner on the  11th of September 2001 after they were struck by two aircrafts.

A model of a gravity-driven collapse of a tall building has been  proposed by Bažant et al. 
We apply this model to the collapse of the North Tower to determine
the energy dissipation of buckling columns per storey during the collapse. This has already been done by 
Bažant et al.~for the first three seconds. 
Using video record data we extend this time range to over 9 seconds. 
Our findings are 250\,MJ during the first 4.6 seconds.
In the time interval between 4.6 and 7.7 seconds after collapse initiation
we find an additional energy dissipation per storey of 2500\,MJ.
Because the steel columns increase in strength towards the ground 
this value corresponds to a value of 2000\,MJ for the storeys in the aircraft impact zone.
After 7.7 seconds the value reduces to the value that corresponds to 
the value during the first 4.6 seconds.

These results have two possible interpretations:
\begin{enumerate}
\item
If due to the building design (column strength, shape etc.) the energy dissipation per storey cannot reach the high values 
which we observed, then the collapse cannot be described by the gravity-driven collapse model.

\item
If the collapse is described correctly by the gravity-driven collapse model,
then we fond direct evidence that the collapse mechanism did not follow the same 
pattern during the whole of the collapse. The possible amount of energy dissipation 
was reduced by an order of magnitude during two long time time intervals.
\end{enumerate}
In both cases there is no a priori reason to justify the sometimes expressed 
belief that 
the collapse was inevitable even after the falling top section had gained 
a significant amount of momentum.  
In fact, if the amount of energy dissipation had stayed only a little longer 
on the high level, then a gravity-driven collapse would have arrested.

Note that (1) implies that if in principle the gravity-driven collapse model describes gravity-driven 
collapses of tall buildings, then the collapse was not gravity-driven.

\end{abstract}

\maketitle


\tableofcontents

\sloppy

\section{Introduction}
\noindent
\subsection{The Case}
On the 11th of September 2001 three major buildings collapsed in New York City.
They were part of the World Trade Center complex which consisted of seven buildings
overall. 
The focus in this paper lies on one of the three. It was called 
the North Tower of the World Trade Center.
At the  time it was built it was the tallest building in the  world with a height of $417\m$,
 110 storeys  and a huge  antenna on top. In the  morning of the 11th of September 2001
 it was struck by an aircraft. The fuselage of the aircraft impacted on the height of the 96th storey
roughly 50\m below the top. The whole building collapsed 102 minutes later \cite[Ch.\,2]{NIST1}.

\subsection{An Attempt for Explanation: The Gravity-Driven Collapse}

An American government agency, the National Institute of Standards and Technology, 
issued a report in 2005 that tried to explain how the 
collapse initiated \cite{NIST1}. However, they did not 
target the question how the collapse progressed. 
Two years later, in 2007, a  
model was proposed by Bažant and Verdure   
that describes the  collapse of the North Tower as
a gravity-driven progressive collapse \cite{BaVe07}.

Therein the collapsing building 
is modelled by three distinct parts which are:
1. The initial top \red{segment} that sat above the first failing floor (this \red{segment} keeps its height until the crushing front hits the ground).
2. The \red{segment} below the top \red{segment} which is compacted from its original undamaged 
size and
moving with the same velocity as the top \red{segment} (the height of this \red{segment} 
is growing in time).
3. The resting, still undamaged \red{segment} below these two (the height of this \red{segment} is reducing).

 \begin{figure}[]
	\includegraphics[scale=0.4, angle=0]{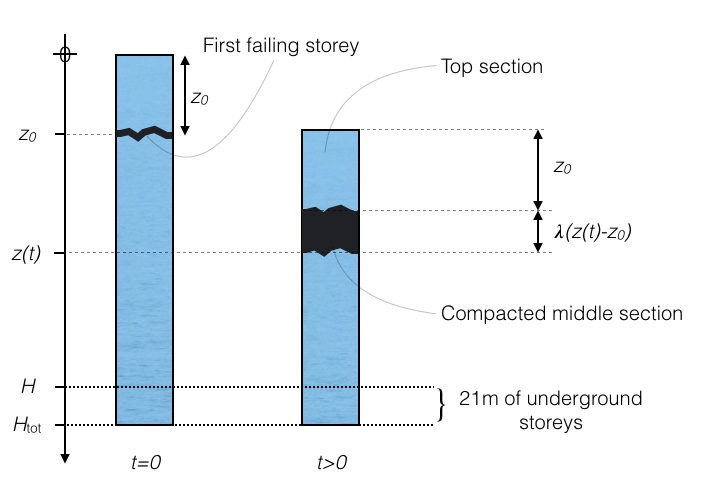}
	\caption{Schematic illustration of the assumed collapse of the North Tower.}
	\label{Schema}
\end{figure}

 During the course of the collapses the initial top \red{segment} 
 stays undestroyed and the  height of the falling \red{segment} (the top and the middle \red{segment} together)
 is strictly increasing until the crushing front reaches the ground.
Then the top \red{segment} is destroyed. 
It should be emphasised  that this behaviour—the undestroyed top \red{segment}—is not 
a choice of model parameters but a consequence of the underlying Newtonian equation of motions. 
 The argument for this conclusion is sketched in \cite[Appendix]{BBGL08} 
 where a two-sided front propagation is computed.
 The upward directed crushing front stops within a fraction of a second
 after having propagated an extremely short distance only.\footnote
 {
 Unfortunately, the authors do not comment why they include 
 the term $\frac{m_2}{2}\dot x$ to the momentum of the top \red{segment}  
 in (33) of \cite{BBGL08} or (4) of \cite{BaLe08}. 
 A term like this could be considered  in (32) 
 as a result of the first colliding storeys. 
 In (32) this term would avoid a vanishing mass of the 
 compacted layer at time 0 if the initial conditions $x(0)=z(0)$
 are used. (A non-vanishing mass is needed to solve the equation for $\ddot z$.)  
 }
So up to this initial and neglectable decay the height of the falling \red{segment} must increase.
 
 The amount of how much the falling \red{segment} is growing in height is specified by how much 
 the middle \red{segment} is compacted. 
   When the crushing front reaches a storey,  
 the ratio  of the full height of that storey divided by  the height after the crushing front 
 has passed is called the compaction parameter $\lambda\in[0,1]$.
In \cite{BaVe07} it is assumed that all storeys are compacted to the same height,
and a value of $\lambda=0.18$ is used.
So if $\kappa_{\rm out}\in [0,1]$ is the parameter 
that specifies the fraction of material that is spit outwards during the collapse at the crushing front, 
then
\begin{eqnarray}
\lambda =(1-\kappa_{\rm out})\nicefrac{V_1}{V_0},
\end{eqnarray}
where $V_0$ is the initial volume of the tower,
$V_1$ is the volume of the compacted rubble pile of the tower.
In the numerical analysis of \cite{BaVe07} $\kappa_{\rm out}=0$ is used, 
in \cite{BBGL08} a value of $\kappa_{\rm out}=0.2$
is considered to be reasonable. The actual value of $\kappa_{\rm out}$ 
does effect the downward movement only gradually (cp.~Figure~\ref{Fit-Sauret}).
In any case, if the crushing front has propagated a certain distance, then 
the height of the falling \red{segment} has increased by $\lambda$ times this distance.

Now let us fix a coordinate system which is pointing 
downwards to the ground and whose origin has a fixed elevation above concourse level,
namely the elevation of the initial undestroyed tower top (cp.~Figure~\ref{Schema}). 
Let $z_0>0$ be the position of the 
storey that collapsed first at the time of collapse initiation ($t=0$), 
i.\,e.~$z_0$ is the height of the undestroyed top \red{segment}.
If $z(t)> z_0$ is the position of the crushing front at time $t>0$,
then $z(t) - \lambda(z(t)-z_0)-z_0= (1-\lambda)(z(t)-z_0)$ is the position of the roof top at time $t$, 
and its time derivative $(1-\lambda)\dot z(t)$ is the downward velocity of both the top and the middle \red{segment}.
Therefore the total momentum of the falling two \red{segments} is
given by $p(t)=m\big(z(t)\big)\,  (1-\lambda)\, \dot z(t)$,
where
\begin{eqnarray}
\label{AggMass}
m(z):=\int_{0}^{z_0}\mu(x)\,dx+(1-\kappa_{\rm out})\int_{z_0}^z\mu(x)\,dx
\end{eqnarray}
describes the accumulated mass of the two moving \red{segment}s. 
 $\mu(\cdot)$ is the mass height-density   
of the undestroyed tower. 
 
Then the equation of motion—which is called Crush-Down Equation in \cite{BaVe07}— that is
 valid until the crushing front reaches the ground is given by
\begin{eqnarray}
\label{crushdown}
\frac{d}{dt}\Big(m\big(z(t)\big)(1-\lambda)  \dot z(t)\Big)=m\big(z(t)\big)g-F\big(z(t)\big),
\end{eqnarray}
where $F(\cdot)>0$ is the  upward resistance force
due to column buckling, and $g$, evidently, is the acceleration of gravity in New York City.
\orange{Note here that the term column force always refers the upward directed force 
exerted by the intact building against the falling upper segments during the actual  collapse.
This quantity is related but not identical to the column strength itself, which 
is an abstract quantity that can be measured under controlled conditions 
in a corresponding experiment. (See Section \ref{Magnitude} below for a further discussion.) }

To model the aircraft impact damage and the fire damage of the tower 
let $\chi(z)\in [0,1]$ be the
parameter which specifies how much the columns are weakened at $z$.
$\chi(z)=1$ means full support.
So the upward force $F$ is the product 
$F=\chi\cdot F_0$, where $F_0$ describes the undamaged column force.
For our numerical analysis we will use
\begin{eqnarray}
\label{damagefunction}
\chi(z)&=&
\begin{cases}
0.5,	&\text{for }z\in[z_0, z_0+h )\quad \text{(first failing storey),}\\
0.9,	&\text{for }z\in[z_0+ h,z_0+4\,h)\quad \text{(impact zone, cp. Fig.~\ref{p22-1}),}\\
1,	&\text{for }z\ge z_0+ 4\,h\quad \text{(intact building).}
\end{cases}
\end{eqnarray}
\orange{Note that the concrete choice of $\chi$ is only of minor importance, as it only effects the solution of the 
Crush-Down Equation in the very beginning. It is merely a way to trigger the propagation of the Crush-Down Equation 
from a resting upper building segment. (We don't make any statements about what happened prior to the collapse or about the 
mechanism that led to a reduction of the column force.)}

The shapes of $\mu$ and $F_0$ are specified in
\cite[Fig.\,2(a)]{BBGL08} essentially as  piece-wise linear 
functions, where the slope of the linear increasing part of $F_0$
is chosen proportional to the (increasing) 
cross-sections of the columns:\footnote{
It is mentioned on pages 895, 896 of \cite{BBGL08} that the transition into the linear 
increasing part should happen at the 81 storey for $\mu(\cdot)$
and $F(\cdot)$. Therefore the term $29\,h$ appears. The building had 110 storeys.  
}
\begin{eqnarray}
\label{MyFirstMu}
\mu(z)&=&\mu_0\cdot
\begin{cases}
1,	&\text{for }z \in[z_0, 29\, h) ,\\
1+0.43\cdot\frac{z-29\, h}{H-29\, h},	&\text{for }z\ge 29\,h,
\end{cases}
\\
F_0(z)&=& \frac{W_{}}{h}\cdot
\begin{cases}
1,	&\text{for }z \in [z_0,29\, h) ,\\
1+6\cdot\frac{z-29\, h}{H-29\, h},	&\text{for }z\ge29\,h,
\end{cases}
\end{eqnarray}
where $\mu_0=\mu(z_0)$ is a constant, $h$ is the height of one storey, $H$ the height of the tower, and
$W$ is the maximal energy absorption capacity of the buckling columns per storey
at the height of the aircraft impact. 
\gold{
In other words: The quantity $W$ is the energy dissipation per storey 
that is needed to crush the columns of building in one storey, i.\,e. up to the constant factor $h$, the quantity 
$W$ is the average upward resistance force during the collapse.
}

\begin{figure}[]
	\includegraphics[scale=0.24, angle=0]{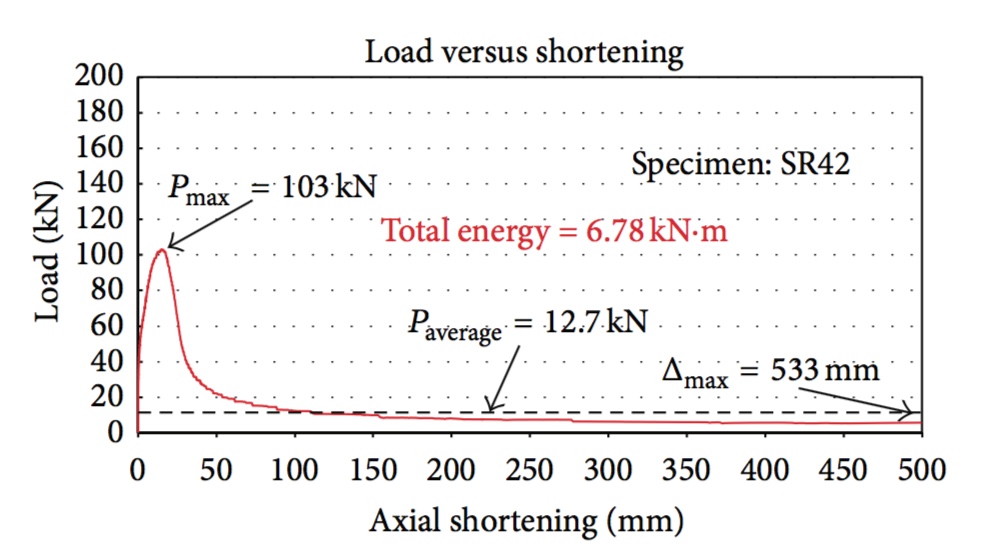}
	\includegraphics[scale=0.15, angle=0]{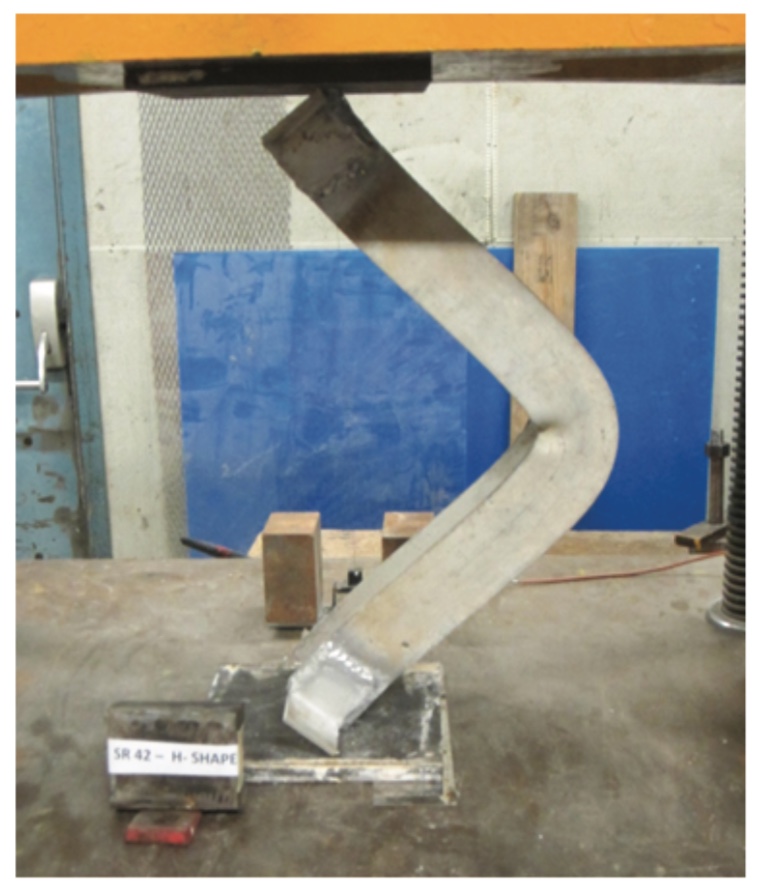}
	\caption{\red{Axial load-displacement plot for an H-profile column taken from \cite[Figure~6]{KS14}
	(to the right a photograph of the experiment).  }}
	\label{Load-Fig}
\end{figure}

Of course, in a realistic scenario the force $F_0$ is not piece-wise linear,
and the model will give unphysical solutions
if the parameters are close to collapse but in reality still 
stable. 
One could \red{simply} add to  $F_0$  a periodic function with period $h$
and vanishing integral to enhance the model to get 
rid of these unphysical solutions.
\red{
Over one period (i.\,e. one storey) the actual shape of such a function could be chosen to be of the shape of the red function in Figure~\ref{Load-Fig}. 
It shows the axial load-displacement curve of an actual H-profile column that has been buckled under controlled conditions.}
\orange{Note here that the function in Figure~\ref{Load-Fig} should only give a qualitative idea of how $F_0$  might be modified over one storey. The situation during the actual collapse is more involved: The columns are fixed, not all columns buckle simultaneously, some might brake off from their base etc. We cannot say much about the concrete collapse mechanism and take it as a black box.}
However, \emph{if} collapse occurs, then  only the average over 
one storey is the energetically relevant quantity.  In this case 
$F_0$ should be regarded as the average upward force, and the 
error \red{of the solution computed without the periodicity of the upward force} is tiny.

\red{
Similarly, the mass density $\mu$ is the average of the actual mass density of the tower.
}

\subsection{Initial Conditions}
\red
{
Note that the initial conditions for our investigation of the Crush-Down Equation (\ref{crushdown}) are
 $z(0)=z_0$ and $\dot z(0)=0$, i.\,e. no downward velocity at time $t=0$. 
Nonetheless the evolution of the collapse will start if the value of the
damage function $\chi$ is low enough which is the case for the values specified in equation (\ref{damagefunction}).
 }

\subsection{The Magnitude of Energy Dissipation}
\label{Magnitude}
The main goal of this paper is to determine the quantity $W$ under the premiss 
of a progressive floor collapse of the North Tower as described by the Crush-Down Equation.
\red{For the sake of completeness, let us recapitulate some of the ongoing debate from the literature and clarify our goal:}

In \cite{BaZh02} and \cite{BBGL08} a maximal possible value of
$W=500\,\rm \,\rm MJ$ is mentioned, which is based on 
computations for a three-hinge 
buckling scenario. Yet meanwhile
Korol and Sivakumaran have made empirical 
studies of buckling columns \red{(cp.~Fig.~\ref{Load-Fig})}, which indicate that 
this value should be about 3 to 4 times bigger \cite{KS14}. 
In  \cite{BaLe16} it is suggested 
that this value again should be corrected by a factor 
of $\nicefrac{2}{3}$. Taking these considerations seriously,
a value of $W=1000\,\rm MJ$
up to  $W=1300\,\rm MJ$ or maybe even more 
might be considered as realistic if the collapse mechanism 
is is based on the three-hinge buckling scenario. 
Note that in \cite{JSS13} a value of even
$2700\,\rm MJ$ is proposed.

It should be emphasised that these values do not
take any empirical data into account that are based on observations
from the actual collapsing tower. 

\red{The objective of this paper is to obtain more refined empirical data
form the actual collapse (Section \ref{WTClength}). Indeed, u}sing a larger pool of video record data we shall extend the empirical range 
to over 9 seconds in total.
We observe a slightly smaller value of about 250\,MJ 
for the time period of $4.6\sec$. 
However, we find that this 
value cannot stand the empirical data in the three second time interval 
between $4.6\sec$ and $7.7\sec$, where
we shall find an additional value of more 
than $2500\,\rm MJ$ of energy dissipation \red{of the buckling columns} per storey.
Taking the increasing strength of the columns into account, 
this value corresponds to
a value of more than $W=2000\,\rm MJ$ at impact level.

This amount of energy dissipation would—if it lasted longer—arrest the fall within the next 10 metres. 
However, after that period of three seconds the energy dissipation per storey reduces again
to the initial low value (relative to the value at impact level).

We do not speculate how it was possible that the resisting structure rises and decays by an order of magnitude,
 but do point out that this was the case. 
A thorough investigation of the
collapse mechanism needs to be done 
in order to understand how such an extreme 
difference  of energy dissipation over long time intervals 
was possible.
  
We then conclude in Section \ref{Subsec-Madonna} that there is no a priori reason 
that one should unconditionally assume that the columns 
of the building were designed too weak  to arrest the fall 
even after the falling top \red{segment} had gained a 
significant amount of momentum.

\subsection{The Modified Model}
 
 In  \cite{BBGL08} the derived Crush-Down Equation is  
 modified on the left-hand side as well as on the right-hand side.
 Let us discuss and clarify these modifications.
We start with the left-hand side.
\begin{itemize}
\item[(lhs 1)]
The compaction parameter is supposed to increase 
proportionally with $\mu(\cdot)$. I.\,e.~instead of assuming that 
every storey is compacted to the same height, it is assumed that 
every storey is compacted to the same density.
We do not feel convinced that this necessarily more realistic, because 
it seems reasonable to expect that during the collapse 
the lower storeys are compacted to 
a higher density than the storeys above.
In any case, this is only a tiny modification,
and for simplicity we will ignore it and 
take $\lambda=const$ in what follows.
\item[(lhs 2)]
 The velocity profile of the middle \red{segment} is supposed to be non trivial.
It is assumed to vary linearly
from the top of the middle \red{segment} down to the crushing front. 
However, this modification is not done accurately in 
\cite{BBGL08} for the following  reasons:
\begin{enumerate}
\item[(a)]
If the velocity profile is non trivial, then conservation of 
mass implies that the density of the compacted \red{segment} is also 
varying. Yet in \cite{BBGL08} it is assumed that the density is constant.
\item[(b)]
The linear velocity profile of \cite{BBGL08} is assumed to vary between 
the velocity of the top \red{segment} (at the top of the compacted layer) and the velocity of the 
crushing front (at the bottom of the compacted layer). This is an extremely unphysical assumption, because 
the latter velocity is bigger than 
the first one. Realistically, the velocity at the bottom of the compacted layer 
should be lower than the velocity at the top.
The velocity of the crushing front should not be regarded as 
the velocity of any mass-bearing instance, but as a quantity that 
describes the change of the geometry of the crushing building.
\end{enumerate}
\end{itemize}
The interested reader is advised to have a look at
\cite{Schn17b}, where detailed account of how
to deal with non-trivial velocity profiles in the compacted \red{segment} is given.
Therein a version of the Crush-Down Equation for
a rather general class of non-trivial velocity profiles is derived
for both cases $\lambda=const$ and $\lambda\sim\mu$. 
The result is that the modified left-hand side of the Crush-Down Equation in \cite{BBGL08}
is not only based on unphysical assumptions, but also that the resulting modification
has the wrong sign, and under realistic assumptions 
its absolute value is far too big. 
The wrong sign and the wrong absolut value 
also changes the  solution of the Crush-Down 
equation in the wrong way \cite[Figure~3]{Schn17b}.

In any case these adjustments are small, 
and for simplicity we ignore these technicalities here: We 
do not make any changes on the left-hand side of the 
Crush-Down Equation.

Let us now turn to the modifications on the right-hand side.
The upward resistance force—which in (\ref{crushdown})
is supposed to be the force due to column buckling only—is 
completed by three other terms. 
They originate in
the pulverisation of the concrete floor slabs, 
the kinetic energy of the ejected air in the squeezed storeys  
and the kinetic energy of the solid ejected material ($\kappa_{\rm out}\not= 0$).    

\begin{itemize}
\item[(rhs 1)]
The term due to ejection of solid material from the tower is 
derived from the assumption that a certain fraction $\kappa_e\in[0,1]$
of all the outwards-thrown material is kicked out 
at the crushing front with the velocity 
of the falling \red{segment}, and the other fraction 
of material has vanishing velocity.
This implies that the term
 \begin{eqnarray*}
&&\underbrace{\frac{1}{2}\kappa_e\kappa_{\rm out}\, (1-\lambda)^2}\,\mu(z)\,\dot z^2.\\\nonumber
&&\qquad\quad\ =: \alpha
\end{eqnarray*}
should be added to the upward Force $F$. 
Note that in \cite{BBGL08} the factor $(1-\lambda)^2$ does not
appear as the bottom of the compacted layer is assumed to move with the 
velocity of the crushing front. 
 Of course, this factor can be suppressed by rescaling $\kappa_e$.   
A value of $\kappa_e=0.2$ is used in \cite{BBGL08}.
 \item[(rhs 2)]
 Once the crushing front has passed, the air inside  
 a crushed storey got ejected. 
 This causes an additional term
\begin{eqnarray*}
\beta \cdot \dot z^2
\end{eqnarray*}
that should be added to the upward force. 
Here no term $(1-\lambda)^2$ appears, as 
the ejection of air is due to the geometric changes
of the collapsing building,
which happen with velocity $\dot z$ (and not with 
the velocity of the falling \red{segment}).
For the precise structure of 
$\beta$ see \cite[p.\,897]{BBGL08}, where 
a numerical range of $\beta$ from 
approximately $40\cdot10^3\,\nicefrac{\rm kg}{\rm m}$
to  $100\cdot10^3\,\nicefrac{\rm kg}{\rm m}$ is 
derived.
The higher values of $\beta$ seem to be rather artificial, 
as the air in the building might also escape through the 
elevator shafts and through the brocken floor slabs
 of a collapsing storey. This is has not been 
taken into account in \cite{BBGL08}.
We will use a value of $\beta=50\cdot10^3\,\nicefrac{\rm kg}{\rm m}$.

 \item[(rhs 3)] For the 
pulverisation of concrete another term is
brought into the Crush-Down Equation:
\begin{eqnarray*}
&&\underbrace{\gamma \cdot \frac{m_c}{2h}} \cdot \dot z^2,\\\nonumber
&&\quad=:\beta'
\end{eqnarray*}
where $\gamma$ is a constant and $m_c$ is the mass of the concrete floor slabs.
The numerical value used in \cite{BBGL08}
is $\beta'=55\cdot10^3\,\nicefrac{\rm kg}{\rm m}$.

No clear explanation is given in \cite{BBGL08} why this term should occur:
The Crush-Down Equation expresses the change of momentum 
under a continuous series of collisions. The total momentum after impact is not effected if the
colliding objects break into pieces or if they stay intact.

In any case, this term has the same structure ($const\cdot\dot z^2$)  as the term in (rhs 2), so
 the general structure of the Crush-Down Equation does not change if  the term is included or not,
and we will only refer to  $\beta$.
\end{itemize}
To summarise the modifications on the right hand side: 
A term
of the form
 $-(\alpha\, \mu(z) +\beta)\,\dot z^2$
should be added.

\subsection{The Downward Movement (Part 1)}
For the numerical analysis let us transform the 
Crush-Down Equation, which is a 1-dimensional differential equation of
2nd order, into its corresponding 2-dimensional equation of 1st order.
Firstly, it can be 
rewritten as
\begin{eqnarray}
\label{crushdownMod}
\ddot z=\phi(z)-\psi(z)\,\dot z^2, 
\end{eqnarray}
where 
\begin{eqnarray}
\phi(z)&=&\frac{g}{(1-\lambda)}-\frac{F(z)}{(1-\lambda)m(z)},\\\nonumber\\\nonumber
\psi(z)&=&\frac{(1-\kappa_{\rm out})\mu(z)}{m(z)} + \frac{\alpha\,\mu(z)+\beta}{(1-\lambda)\,m(z)}.
\end{eqnarray}
Secondly, if $(z,u)\mapsto X(z,u)$ is the vector field given by
\begin{eqnarray}
X(z,u):=\left(
\begin{array}{c}
u\\
\phi(z)-\psi(z)\,u^2 
\end{array}
\right),
\end{eqnarray}
 we shall consider the equation
$\frac{d}{dt}(z,u)=X(z,u)$, which is equivalent to the original Crush-Down Equation.
To analyse 
this equation numerically
we use the open source computer algebra system
Maxima (wxMaxima 16.04.0, \cite{Maxi}) 
that is equipped with a pre-implementation of the Runge-Kutta algorithm. 
The source code is given in Appendix~\ref{Numerics}.
Figure~\ref{BazantSolution}
shows the height of the tower top 
as a function of time as derived from the 
Crush-Down Equation.
The solutions in the left diagram are computed 
for the following choice of parameters:
\begin{eqnarray}
\label{Parameters}
\\\nonumber
\begin{array}{r c l r c l r c l l l}
\beta&=&0.05\cdot 10^6\,\nicefrac{\rm kg}{\rm m},&H&=&417\m,&z_0&=&46\m&\text{(cp.~Section~\ref{SecES}),}\\
\kappa_{\rm e}&=&0.2,&h&=&3.8\m,&
 \dot z(0)&=&0,&\\
 \mu_0&=&0.6\cdot 10^6\,\nicefrac{\rm kg}{\rm\!m},& g&=&9.8\,\nicefrac{\m}{\sec^2}.&\\
\end{array}
\end{eqnarray}
For comparison we have included the right diagram with a total mass of the tower increased  by 50\%.
For  $\mu_0=0.57\cdot 10^6\,\nicefrac{\rm kg}{\rm m}$ the total mass of the tower (including 21\m of 
underground storeys \cite[p.\,19]{NIST1-1}) is 288,000\,t. This  value has been 
estimated meticulously  in  \cite{Uri07}.
In \cite{BBGL08} a value of 500,000\,t is stated without reference,
which would give  $\mu_0=0.98\cdot 10^6\,\nicefrac{\rm kg}{\rm\!m}$.

\begin{figure}
	\includegraphics[scale=0.52, angle=0]{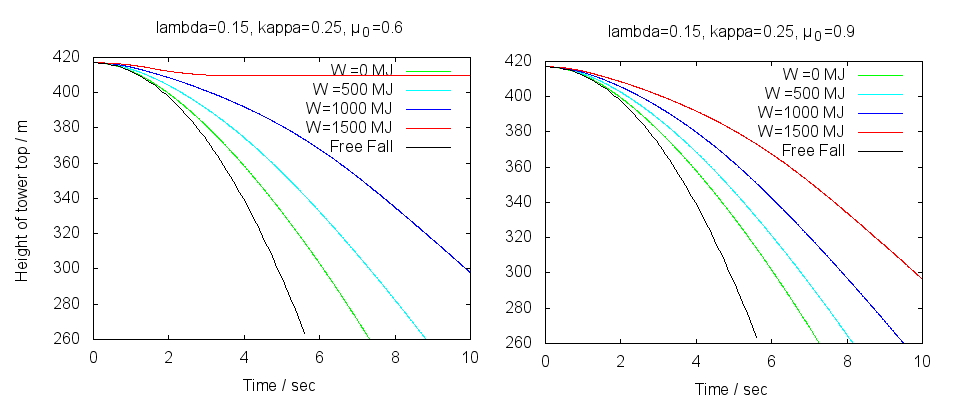}
	\caption{Solutions of the Crush-Down Equation (converted into the
	elevation of the roof by $z\mapsto H-(1-\lambda)(z-z_0)$) for two different masses of the tower. 
	The left diagram was computed for $\mu_0=0.6\cdot10^6\,\nicefrac{\rm kg}{\!\rm m}$, the right one 
	for $\mu_0=0.9\cdot10^6\,\nicefrac{\rm kg}{\!\rm m}$.}
	\label{BazantSolution}
\end{figure}

\begin{samepage}
\section{Length Measurements of the North Tower's Collapse}
\label{WTClength}
\orange{
\noindent
\subsection{Coupling the Model to Empirical Data}
So far we have developed the mathematical framework to describe the collapse of the North Tower. We are now in the position to feed the model with empirical data to extract $W$, the energy dissipation of the squashing building per storey during the collapse. The Crush-Down Equation enables us to recompute the upward force $F_0(\cdot )$ (i. e. its scaling factor $W$)  from the downward movement $t\mapsto z(t)$ of the crushing front. For that we need to determine the downward movement of the crushing front from the actual collapse. This is the contend of the following empirical analysis.
}

\noindent
\subsection{The Idea of Measurement}
Our plan is to analyse
\end{samepage}
 video footage from  different records of the North Tower's Collapse.
 The  sources are a short film documantary 
 by Etienne Sauret called {\sl 24 Hours} \cite{Saur02}
 a History Channel documentary called {\sl The 9/11 Conspiracies: Fact or Fiction}
 \cite{Hist} and some footage from CBS \cite{CBS01} and CNN \cite{CNN}.
 Our goal is to determine the position of the roof
 under the principal assumption that the initially falling top \red{segment} of the building 
 stays undetroyed during the course of the collapse.
 As we have mentioned this principal assumption is 
 a consequence of a gravity-driven collapse.
 
During the first three and a half seconds the top \red{segment} is visible 
in Sauret's video record, which enables a direct measurement of the 
height of the roof.
After the roof disappeared behind the dust cloud 
the antenna is still visible, so we can trace the roof 
by tracing the movement of the antenna.

After the antenna disappeared we can still
make reasonable statements about the 
position of the roof by just estimating 
the crushing front from below.
This is done with the video clips of History Channel and CBS.
The initial height of the top \red{segment} plus 
the height of the compacted \red{segment}  
must be added to the measured lower bound of the 
crushing front to obtain a lower bound for the 
position of the roof.
 
\subsection{Video Analysis Tool and Machine Data}
We do some simple length measurements with the open-source video analysis tool  {\sl{ Tracker}},
Version 4.96 \cite{Tracker}, running on a 2.7\,GHz Intel Core i5 iMac with operating system 
OSX\,10.11.3\,(15D21). It is equipped with an  8\,GB 1600\,MHz DDR3 RAM, and 
an Intel Iris Pro 1536\,MB graphics card.

\subsection{Etienne Sauret}\label{SecES}
We use a sequence of stills  from the short film {\sl 24 Hours} 
shot by Etienne Sauret \cite{Saur02} 
to determine the elevation of the top of the tower 
at three different times after collapse initiation.

Time is always measured relative to the collapse initiation at  $t=0$,
which for us is the first recognisable movement of the north-west corner of the roofline.
In the video copy we use this happens at frame number $934$ (first frame has number 0).
The first visible movement of the antenna is three frames earlier.
The frame rate of the video is 29,97 frames per second, i.\,e.~3 frames in 0.10 seconds,
which means the uncertainty in time is about  0.033\,sec.

Let us now chronologically follow our measurements.

\begin{center}\begin{figure}
	\includegraphics[scale=0.25, angle=0]{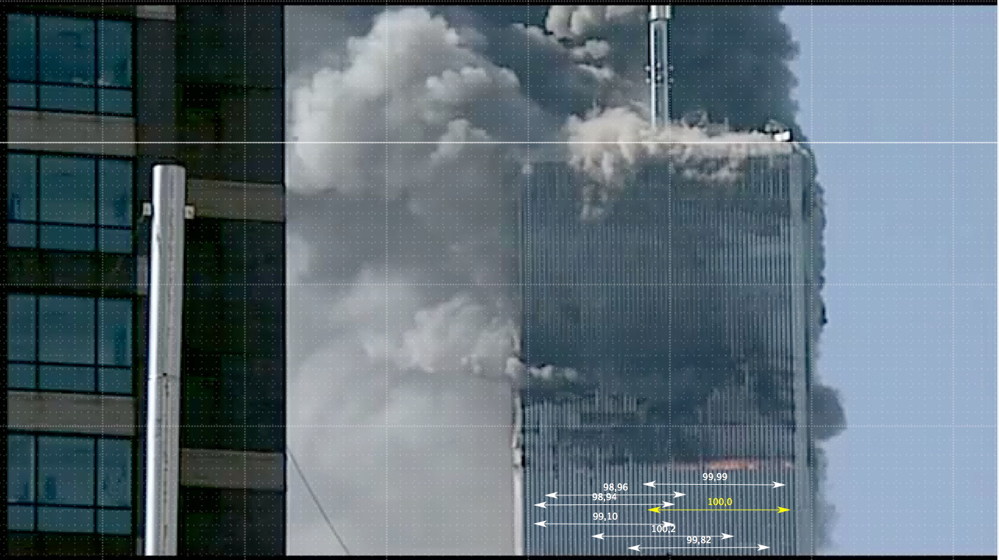}
	\caption{Horizontal calibration at frame 800, $t=-4.47\sec$.
	(For a zoom on the measurements  go to Figure~\ref{800SauretZoom}.)  }
	\label{0800Sauret}
\end{figure}
\end{center}

Figure \ref{0800Sauret} shows frame 800, i.\,e.~$t=-4.47\sec$.
It shows a foreground building in the left. 
All other images are cropped and only show the right part of the actual video image.
The foreground building is 101 Avenue of the Americas  (6th Avenue). The video was shot from
145 Avenue of the Americas which is an 8 storey building (s.\,building description in Figure~\ref{145Avenue}
and Figure~\ref{Google145Ave}), where Sauret's 
film company {\sl Turn of the Century Pictures} was based on the 7th floor \cite{Turn}.
This position gave an almost orthogonal perspective on the north side of the North Tower
with  a distance of $(1,550\pm20)\m$ (Figure \ref{145Avenue}). 
The roof of the North Tower (without its antenna) had an elevation of  
$1,368\,\rm ft =417.0\m$ \cite[p.\,5]{NIST1}. 
The optical center of the camera  is targeting approximately 30\m below the roofline slightly to the east of the building.
After estimating the height of the camera with another $30\m\pm10\m$ we obtain
an upward camera angle of $\arctan(\nicefrac{(417-60\pm10)}{(1550\pm20)})=13^\circ\pm1^\circ$.
The sideward camera angle is estimated by $6^\circ\pm 1^\circ$ (Figure \ref{145Avenue}).
Therefore, measured vertical distances have to be 
scaled up by a factor of $\nicefrac{1}{(\cos((6\pm1)^\circ)\cos((13\pm1)^\circ))}= 1.03\pm0.01$ 
when compared with horizontal distances on the north face of the tower. 
(The camera's angle of view is small and neglected).
Below we shall make an explicit comparison with a known vertical distance.

The horizontal length calibration of the video is done as follows:
Seven times we have measured the distance of thirty columns including the gap 
to the next column,
where we have set one of the distances to a reference scale of 100 units.
Figure~\ref{0800Sauret} shows these measurements. The yellow line
is the reference line. Figure~\ref{800SauretZoom}  
shows a zoom on the relevant part  of
the image.
From top to bottom these seven measurements are:
\begin{eqnarray*}
99.99,\quad
98.96,\quad
98.94,\quad
{\bf 100},\quad
99.10,\quad
102.2,\quad
99.82.
\end{eqnarray*}
The mean is 99.86 with a standard deviation of 1.05.
The structural diagrams of the steel segments used in the construction of the tower are shown
in \cite[p.\,25]{NIST1-1}. The width of a segment of  three columns (including the gap to the next
column) is stated as 10\,ft~0\,in, so thirty columns and the gap to the next one had a
width of $100\,\rm ft~0\,in = 30.48\m.$
Therefore we will use
\begin{eqnarray}
(99.86\pm1.05) \text{ reference units} = (30.48\pm0.32)\m.
\end{eqnarray}

Our baseline for vertical length measurements is the horizontal line
touching the north-west corner of the roof.
This is the slightly thickened line of the grid in Figure \ref{0800Sauret}.

 From Sauret's camera perspective 
the north-east corner appears to be approximately 1\m
lower than the north-west corner. 

We should now compare the horizontal calibration with a known vertical length.
This is done by the measurement in Figure~\ref{800SauretVertCalibration}.
The white line indicates the horizontal calibration of $30.48\m$.
The yellow measurement from the baseline to the red line  gives
a vertical distance of $(1.03\pm0.01)\cdot (60.9\pm 0.6)\m=(62.7\pm 1.3)\m$.
We can identify the red line with the long white line in Figure~\ref{p35-1-5A},
which itself can be identified with the 95th floor (Figure~\ref{p22-1}). 
According to the  structural drawings of the tower, 
the distance from the roofline 
to the 95th floor was $90\,\rm ft~1\,in=63.73\m$ (Figure~\ref{p18-1-1}).
So the deviation is within our range of precision and we can proceed.

Three meters below the roofline the visible end of the steel columns 
appears as a transition from the lighter roof to the darker lower side of the building.
We refer to this line as the `bottom of the roof'. It is sometimes easier to identify than 
the roof itself.
Figure~\ref{934Sauret} shows the collapse initiation at frame 934.
We measure 10\m away from the  corners
the position of the bottom of the roof. We find it
$1.03\cdot 4.2\m=4.3\m$ and $1.03\cdot3.0\m=3.1\m$  below the baseline.
So in the middle we have a  distance of 3.7\m to the baseline, i.\,e.~in 
the middle the top of the roof has a distance of $0.7\m$ to the baseline

Figure~\ref{957Sauret} shows frame 957, $t=0.77\sec$.
The middle one of the three yellow arrows points from the top of the roof
to where one might think the collapse initiated at the north-west corner of the building. 
This is the middle of the lighter part of the appearing dust cloud. It has a length of $1.03\cdot 44.76\m= 46.1\m$. 
The arrow in the left indicates a part of the perimeter columns which move simultaneously
with the top \red{segment}: As far as one can say, at this stage the crushing front is no clean horizontal line.
The yellow line to the right has a height of $z_0=1.03\cdot77.7\m=80\m$, which 
is the height that is  used in \cite[Fig.\,6]{BaVe07}  for the initially falling block. Clearly, this is 
an overestimation of that height. 
It has already been pointed out in \cite{JSS13} that the
mass of the falling block has been overestimated in \cite{BaZh02,BaVe07,BBGL08}.
The wrong height assumption is probably the origin of this error, because
the values match the mass distribution functions given in \cite[Fig.\,6]{BaVe07}. 
Note that in \cite[p.\,151]{NIST1} it is mentioned that the collapse initiated
at the 98th floor. According to the structural drawings (Figure~\ref{p18-1-1}) the 12 storeys 
above had a height of $162\,\rm ft~1\,in=49.6\m$.
We will use therefore use $z_0=46\m$ as a {\emph {lower bound}} for the height of the initially 
falling block.

The red line in Figure~\ref{983Sauret} has an angle of $2.4^\circ$.
It shows  frame 983, $t=1.64\sec$.
We find the top of the roof at $1.03\cdot (12.7-3) \m =10.0\m$ 
below the baseline (the green measurement line ). 

The antenna had clearly recognisable \red{segment}s that appear white and dark from the front perspective.
At frame 1024, $t=3.00\sec$, the bottom of a white part is visible at the top of the video.
The measured distance between the lowest point of the white part of antenna and 
the bottom of the roof is  $60.4\m$. 
This is the  light blue line in Figure~\ref{1024Sauret}.
The antenna has an angle of approx. $2^\circ$ to the east at this time.
(The antenna's angle to the south reaches a value of $8^\circ$ before it is not visible any more.
See below.)

The red line in Figure~\ref{1030Sauret} has an angle of $3.7^\circ$.
It shows  frame 1030, $t=3.20\sec$.
We find the top of the roof at $1.03\cdot (36.8-3) \m =34.8\m$ 
below the baseline 
(the green measurement line ). 
The roof is still visible until couple of frames later 
but the contour is getting weaker as 
it gradually disappears behind the dust cloud.

At frame 1050 the light part of the antenna is completely visible and measured. This is the 
short light blue line showing a measured length of $19.4\m$
in Figure~\ref{1050Sauret}. Together with the lower part 
we find the measured length of these two antenna \red{segment}s to be $79.8\m$.
The eastward angle of the antenna is $5^\circ$.

Figure \ref{1071Sauret} showes frame 1071, $t=4.57\sec$. 
This is the last frame where the top part of the white antenna \red{segment} is still visible.
The light blue line of 79.8\m length indicates the position of the antenna with an 
assumed angle of $9^\circ$ to the east. The distance from the baseline to its lowest point is therefore
$1.03\cdot 70.0\m=72.1\m$.
This is the point where the bottom of the roof is at this time. Of course, we assume here that 
the roof still exists.
The antenna not only tilted eastwards but also southwards. In \cite[p.\,166]{NIST1-6}
an angle of $8^\circ$ is mentioned. 
So there is a small additional correction factor of $\cos(8^\circ)=0.99$,
which gives a decent of $0.99\cdot 1.03\cdot (70.0-3)\m=68.3\m$
for the top of the roof.

If we assume the total elevation of the 
middle of the roof 
to be
$417\m$ at collapse initiation, we  can 
summarise the measurements in the following table,
including appropriate error estimates.
The error estimate for the last value (antenna based)
is bigger as the antenna might not be fixed on the roof, 
as we have mentioned the movement of the antenna 
started little before $t=0$.

\begin{center}
\resizebox{10cm}{!}{
\begin{tabular}{c  c  c  c }
\\
\toprule
Time/sec&Part of roof & 
\begin{tabular}{c} Distance to\\baseline/m \\ \end{tabular}&
\begin{tabular}{c}Elevation over\\ concourse level/m\end{tabular} \\
\midrule
0& top, middle &0.7&$ 417$ \\
1.64& top, middle &10.0&$ 408\pm 2\,\ \ $ \\
3.20& top, middle &34.8&$ 383\pm 2\,\ \ $ \\
4.57&top, middle& 68.3&$349\pm 4\,\ \ $\\
\bottomrule
\\
\end{tabular}
}
\captionof{table}{Results of height measurements.}
\label{FirstResults}
\end{center}
 \begin{center}
\begin{figure}[b]
\hspace{0cm}\includegraphics[scale=0.53,angle =0]{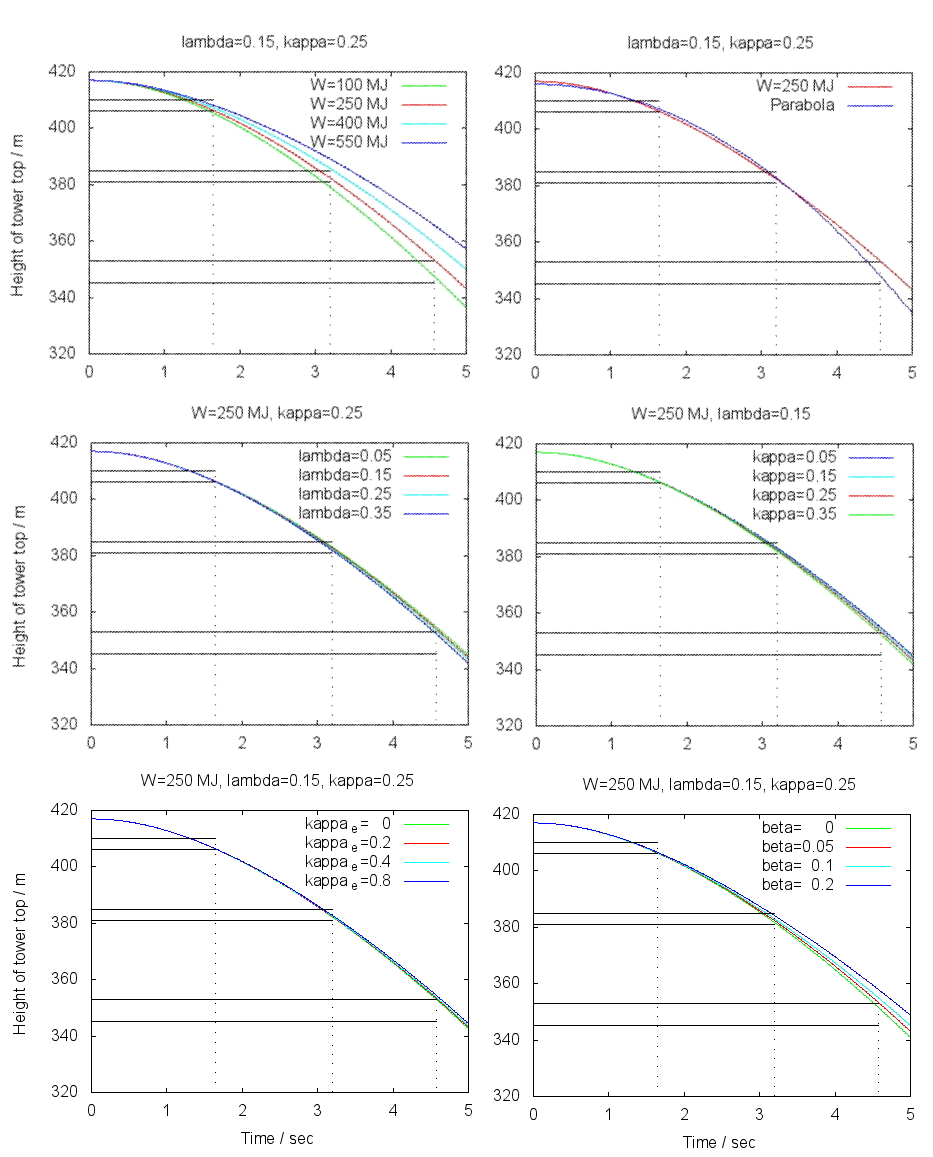}
	\caption{The movement of the roof (top) during the first 5 seconds.}
	\label{Fit-Sauret}
\end{figure}
\end{center}

\subsection{The  Downward Movement (Part 2)}
\label{Part2}
The video material of the Sauret video has already been used in \cite{MaSz09} and \cite{Chan10}
to determine the downward acceleration of the roofline of the North Tower with shorter time intervals
during the first three seconds.
Their basic findings were a movement of the roof with a constant acceleration of
$22.8\,\nicefrac{\rm ft}{\sec^2}=6.95\,\nicefrac{\m}{\sec^2},$ and $6.31\nicefrac{\m}{\sec^2}$, respectively.
To quickly compare our results with these two we do a linear regression 
for a parabola
$t\mapsto\frac{1}{2}\, a\, t^2+b$.
The four data points from the first and third column of Table~\ref{FirstResults} give
\begin{eqnarray}
\label{Parabola}
a=6.46\nicefrac{\m}{\sec^2},\quad
b=1.14\m,\quad
r=0.99988,
\end{eqnarray}
where $r$ is the regression coefficient.

The empirical data from Table~\ref{FirstResults} are illustrated in 
the six diagrams of Figure~\ref{Fit-Sauret} by the horizontal black lines which indicate 
 the error bars. The actual values in the middle are not displayed. 
 The coloured curves are the predicted model curves 
 for the indicated values and for the other parameters 
 as given in (\ref{Parameters}).
 The parabola in the upper right corner is the one derived from (\ref{Parabola}) and displayed for 
 reasons of comparison.

For simplicity we do not give a sophisticated optimisation analysis here,
but based on the printed model curves we take  $W=250\,\rm MJ$, 
$\lambda=0.15$, $\kappa_{\rm out}=0.25$, $\kappa_{\rm e}=0.2$, $\beta=0.05\cdot 10^6\nicefrac{\rm kg}{\!\rm m}$
 as the  result with which we continue to work. 
The precise values for the best fit will not be important for our main result (cp.~Section~\ref{Part3}). 
The red graph in all six diagrams shows this solution.

Note, firstly, that higher values of $\kappa_{\rm out}$ become more and more unrealistic
in a gravity-driven collapse.
Secondly, a higher value of $\beta$ would require a lower value of $W$.\footnote{
The same is true for the erroneous assumption on the velocity profile of the middle \red{segment}
which would require a lower value of $W$ as explained in \cite{Schn17b}.
}
Thirdly, a lower value of $W$ would  also match better if the starting time of the model curve
is put slightly later at $t_{\rm late}=\tau\sec$, $\tau\in[0,0.2]$. This would be a legitimate adjustment, 
 as the model only describes the dynamical aspect of the collapse itself.
It does not model the transition from the stable to the unstable state which takes a finite time interval.
In this respect the value of $W=250\,\rm MJ$ is only an upper bound  of energy dissipation 
for the first $(4.6-\tau)\sec$. This is important later.

\subsection{Comaparing with Prior Obtained Data}
Unfortunately, there is no reference in \cite{BaVe07,BBGL08}
about the video footage that has been used and no indication 
about the starting time for their measurements. So neither we can 
comment on their starting time nor on the accuracy of their 
measurements. 
The result of \cite[p.\,902]{BBGL08} is
an average upward force due to column buckling of $0.1\,\rm GJ/m$. 
This correspond to a value of energy dissipation 
of $W= 380\,\rm MJ$ per storey ($h=3.8\,\rm m$).

\begin{center}\begin{figure}[h]
	\includegraphics[scale=0.8, angle=0]{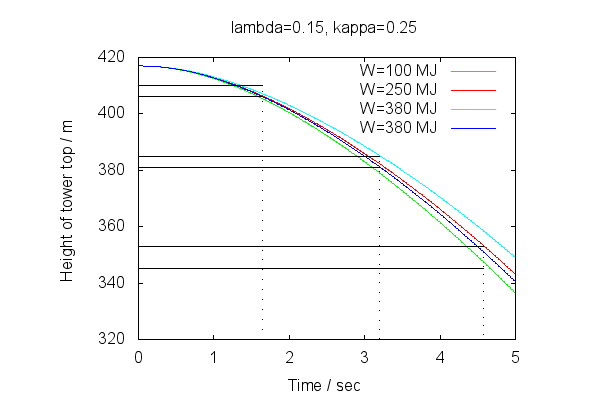}
	\caption{Comparing solutions of the Crush-Down equation for different masses. The dark blue solution is computed for an overall  mass of the tower of
	 500,000 tons. The red, green and light blue solutions are computed for a mass of 300,000 tons.}
	\label{Compare_Schneider_Bazant}
\end{figure}
\end{center}

The discrepancy of our 250\,MJ and the value of 380\,MJ of
 \cite{BBGL08} (cp. \ref{Magnitude})
might be mainly explained due to the different numerical values of the mass of the tower.
In  \cite{BBGL08} a mass of 500,000 tons is used, whereas we use a mass of 
300.000 tons. 
A direct comparison of the two solutions (250\,MJ for 300.000 tons and 
380\,MJ for 500,000 tons)  is given in Figure \ref{Compare_Schneider_Bazant},
where it is observed that their difference is rather small.

\subsection{History Channel}
\label{HistoryC}
We only want to evaluate one still from a documentary aired on History Channel \cite{Hist}.
The frame rate of this video is 59.97 frames per second, so 6 frames correspond to 0.1 seconds.
This footage shows the destruction of the North Tower recorded from West Street
from a north-west ground perspective.
It does not show all of the collapse, as the first few seconds are missing.

\begin{center}\begin{figure}
	\includegraphics[scale=0.19, angle=0]{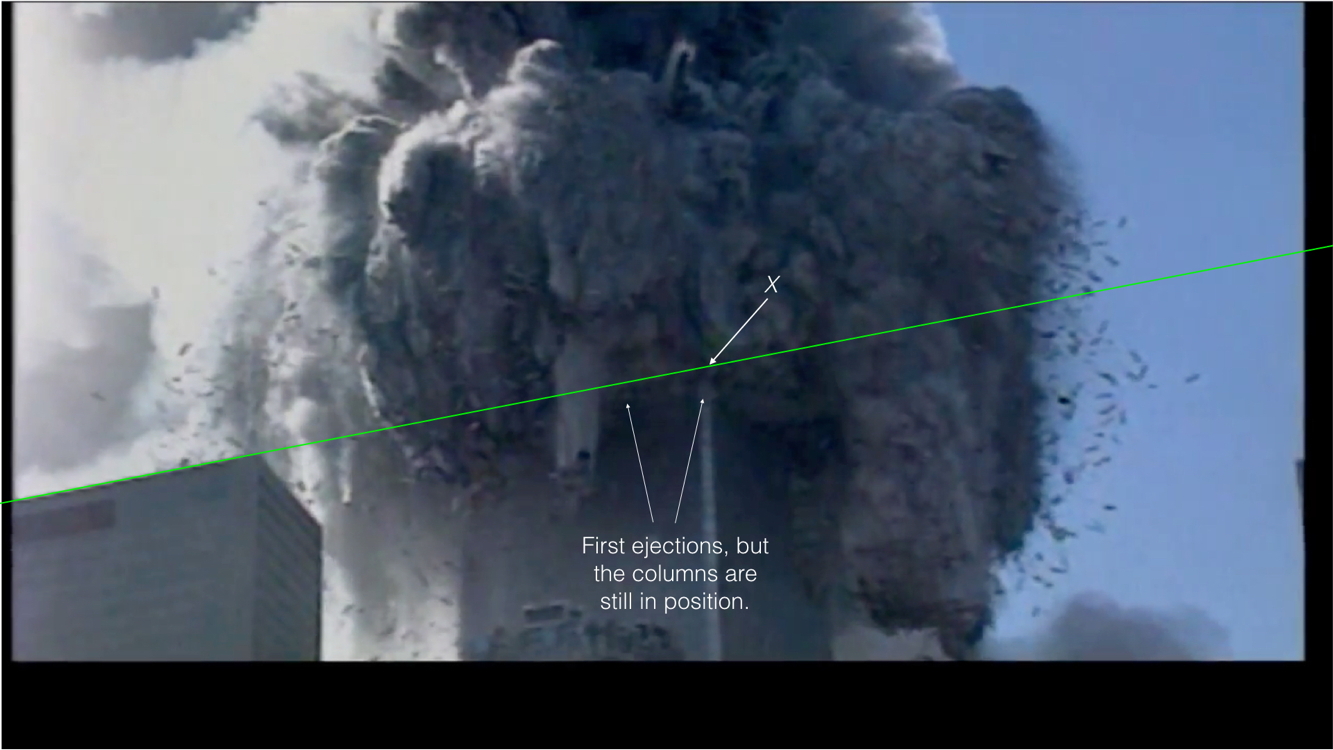}
	\caption{The collapsing tower from West Street, frame 262 of the History Channel clip.}
	\label{262History}
\end{figure}
\end{center}

Figure~\ref{262History} shows the collapsing tower at frame 262. 
It is possible to identify the time of frame 262 with a precision of one frame in Sauret's video.
This is done in Appendix~\ref{AppTime} and the result is
$t=(7.71\pm{0.033})\sec$.

The building to the left is WTC\,7. It 
was one of the three high-rise buildings that collapsed
on the 11th of September 2001.
It had 47 storeys and its roofline had a height of 
$h_1= 610\,\rm ft=185.9\m$ \cite[p.\,5]{NIST1A}.
The green line follows the roofline. 
Once the camera position is known we can determine the height $h_X$ of the 
point $X$ that is  behind the green line right on the corner of the tower.
The camera position is determined in Appendix \ref{HistoryApp}.
If we assume the height of the camera to be $h_0=1.7\m$,\footnote
{
Because the distance $d_0$ is much bigger than $d_{\rm int}$, 
the camera height has no practical influence on the result of (\ref{TheXheight}).
E.\,g.~an additional height of 2\m would reduce the height of the crushing front 
by $2\m\cdot\frac {d_{\rm int}}{d_0-d_{\rm int}}=0.7\m$ only. 
}
then 
the camera was located on West Street, 
in a distance of $d_0=694\m\pm9\m$ away from the north-west corner of the tower 
(Figure~\ref{NYC}).
The distance from the north-west corner of the tower to the intersection of the 
camera line and the projection of the green roof line to the ground
(that's the bold green line in Figures~\ref{NYC}, \ref{NYC3})
is determined to be $d_{\ins}=175\m\pm 2\m$. (That's the red line in Figures~\ref{NYC}, \ref{NYC3}.)
Therefore the point $X$ on the corner of the tower has an elevation of
\begin{eqnarray}
\label{TheXheight}
h_{X}&=& (h_1-h_0)\frac{d_0}{d_0-d_\ins}+h_0\\\nonumber
&=&248\m\pm 2\m.
\end{eqnarray}
Apparently, Figure~\ref{262History} shows that the crushing front is about to reach 
the point $X$. Some amount of dust is already blown outwards
below $X$, but the perimeter columns are still standing without  being affected.
This is agreed in  \cite[p.\,901]{BBGL08}, where it is stated:
\begin{itemize}
\item[]
\textit{“Some critics believe that the bottom of the advancing dust cloud seen in the video represented the crushing front. However, this belief cannot be correct because the compressed air exiting the tower is free to expand in all directions, including the downward direction. This must have caused the dust front to move ahead of the crushing front[\dots]“}
\end{itemize}
In other words the point $X$ is only a \emph{lower bound} for the approaching crushing front. 
However, we shall give an argument in Section~\ref{Part3} that the distance 
between $X$ and the crushing front is probably small.
 
Now recall that the original height of the tower was 417\m and that
the falling upper block had an initial height of at least $46\m.$ 
This means that at the time of frame 262 a 
distance of at most $417\m -46\m-248\m= 123\m$ has been crushed.

If we assume a gravity-driven collapse, then the top 46\m are still undestroyed,
and a falling \red{segment} of height $\lambda\cdot 123\m+46\m$  sits
somewhere
above point $X$ in Figure~\ref{262History}.
We conclude that the roof had a total elevation of at least
$248\m+46\m+\lambda\cdot 123\m$
above concourse level at the time of frame 262.
For a compaction parameter of $\lambda=0.15$ this gives 
an elevation of $312\m$.

Note that if one assumes a bigger height for the initial falling \red{segment} (as in \cite{BaVe07,BBGL08}),
a bigger value of $\lambda$ (as in \cite{BaVe07,BBGL08})
or a bigger height for the crushing front one obtains an even higher
elevation of the roof.

\subsection{CBS}
\begin{center}\begin{figure}
	\includegraphics[scale=0.364, angle=0]{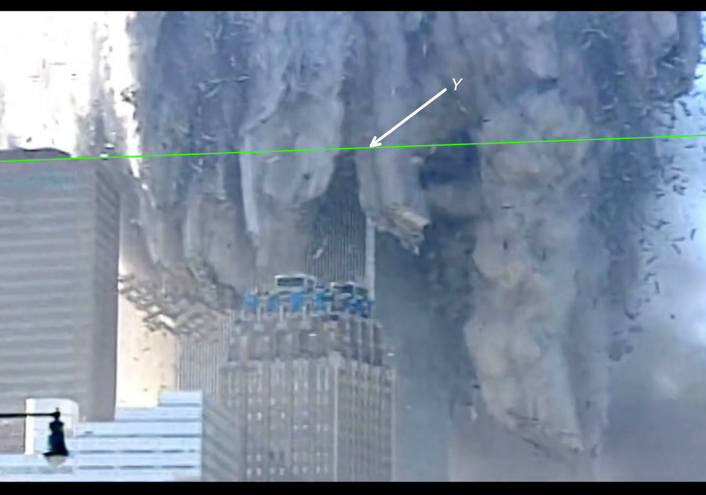}
	\caption{The collapsing tower from West Street, frame 739 of the CBS clip.}
	\label{739CBS}
\end{figure}
\end{center}

\begin{center}\begin{figure}[t]
	\includegraphics[scale=0.3, angle=0]{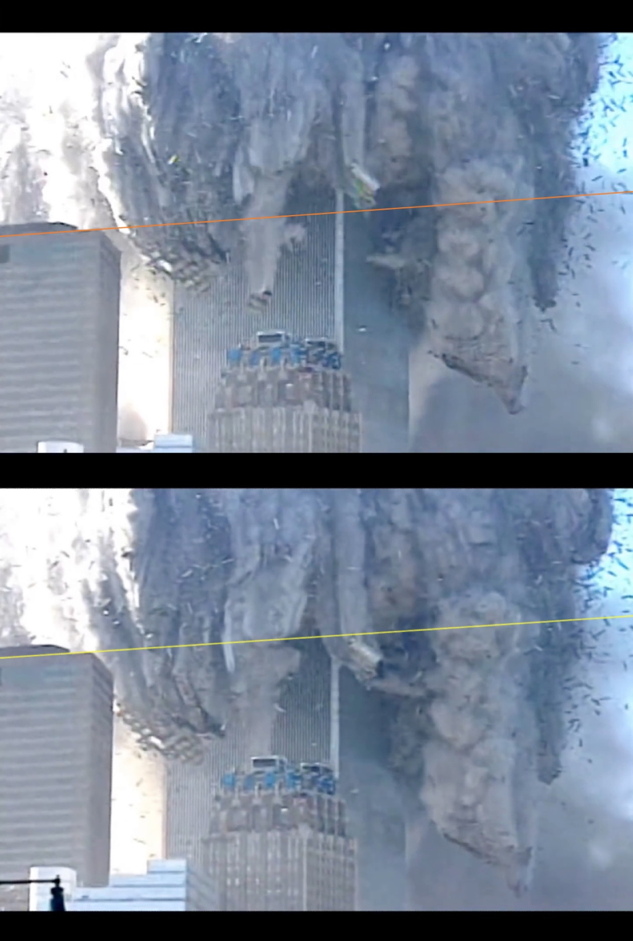}
	\caption{The collapsing tower from West Street, frames 719 (top) and 729 (bottom) of the CBS clip.}
	\label{10and20Frames}
\end{figure}
\end{center}

We do the same routine as for the History Channel clip 
for a clip from CBS. The copy of the film that we use has a frame rate of 25 frames per second \cite{CBS01}. Figure~\ref{739CBS} shows frame 739 of this clip 
at a time of $t=9.25\sec$ after collapses initiation (s.\,Appendix~\ref{CBS-Sync} for synchronising the CBS clip).
It shows that the crushing front 
is about to pass the point $Y$ behind the green line (the  roofline of WTC\,7). 
The falling debris obstructs the view to most of the 
ejected dust from the tower, but going through the actual video clip shows that
the moment is captured correctly in the sense that this is the last moment for which 
 we can conclude that the crushing front is above the point $Y$. For comparison
Figure~\ref{10and20Frames} shows the same clip 10 and 20 frames ($0.40\sec$ and $0.80\sec$) earlier.
 
In Figure~\ref{739-CBS-WTC7} the distance from the 
north west corner to the green line in the direction of the 
CBS camera is measured by the dark blue line. This distance is $d'_{\ins}=(170\pm2)\m$.
In Appendix~\ref{CBS-Posi} we find that the distance between the CBS camera and the north-west corner is 
$d_0'=(1202\pm20)\m$.
So if we again assume a camera height of $h_0=1.7\m$ 
we find another lower bound for the crushing front 
by the height $h_Y$ of the point $Y$ on the tower
at time $t=9.25\sec$:
\begin{eqnarray}
h_{Y}&=& (h_1-h_0)\frac{d'_0}{d'_0-d'_\ins}+h_0\\\nonumber
&=&216\m\pm 1.5\m.
\end{eqnarray}
This estimate implies that a height of not more than
$417\m-46\m-216\m=155\m$ has been compacted.
For $\lambda =0.15$ this gives an elevation of 
$216\m+46\m+\lambda\cdot 155\m= 285\m$.

\subsection{The  Downward Movement (Part 3)}
\label{Part3}
\begin{center}\begin{figure}
	\includegraphics[scale=0.5, angle=0]{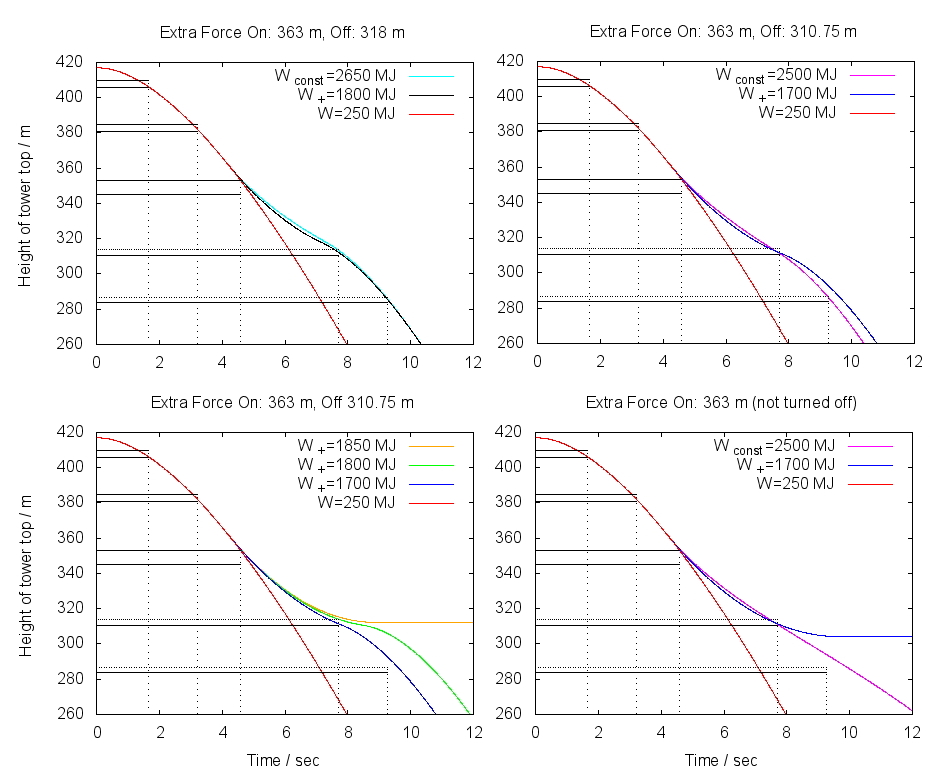}
	\caption{The measured data and the  model curves.}
	\label{Ojemine}
\end{figure}
\end{center}

The error bars for the measured data are shown in Figure~\ref{Ojemine} as before 
by horizontal  black lines. The dotted lines for the two lower 
measurements indicate that this is only a lower bound.

The red graph is identical in all four diagrams and shows the
solution of the Crush-Down Equation 
with the maximal energy dissipation during the first seconds 
 as explained in \ref{Part2}, i.\,e.~$W=250\,\rm MJ$, 
$\lambda=0.15$, $\kappa_{\rm out}=0.25$ and the other
parameters as in (\ref{Parameters}).
Note that the red graph misses the empirical data point at $7.71\sec$ by 40\,m, 
so we detect a major discrepancy here. This discrepancy would be significantly 
bigger for the above discussed value of an energy dissipation of only
 $W=100\,\rm MJ$ 
 per storey.

The graphs in other colours are also solutions of the 
Crush-Down Equation for the same choice of parameters 
except that to the upward force $F_0$ 
an extra upward force is added over a certain interval. 
The interval is indicated above each diagram.
It specifes the position of the roof where the force is
turned on, and the position of the roof where it is turned off again.

Two types of extra forces we have used for the computations: 
$(a)$ A constant force  $F_{\rm const}(z)=W_{\rm const}/h$
and  $(b)$ an extra force $F_+$ that is directly proportional to $F_0$ 
by the factor $W_+/W$, i.\,e. in this case the total upward force is 
again proportional to $F_0$, namely
$F_0+F_{+}=\nicefrac{(W+W_+)}{W}\cdot F_0$.
Therefore the sum $W+W_+$ is the quantity which can be directly
compared to the values discussed in Section~\ref{Magnitude}.
The force $F_+$ is the relevant quantity, as it reflects the 
column strength of the actual building. The discussion of the force 
$F_{\rm const}$ is given for reasons of comparison.

All extra forces are turned on 10\m above the upper 
error bar of the third measurement, i.\,e.~at 363\m in all
diagrams. This  takes the solutions out of the measured  position 
at time $t=4.57\sec$, but we are interested in the minimal extra force 
that must be applied to match the two lower data points. By increasing 
the height where the extra force is turned on, we decrease the value of 
the necessary extra force to reach the lower data points at $t=7.71\sec$ and
$t=9.25\sec$, so this gives 
an error that decreases our result.

Three intervals are considered: Turning off at 318\m, turning off at 311\m and not turing of at all.
The height 311\m is the height of the lower error bar of the measurement at $7.71\sec$.
The extra force is minimised if it is applied until 311\,m.

The magenta graph in the upper right diagram shows that during the 
time interval from $t=4.57\sec$ to $7.71\sec$ an additional  energy of
at least $W_{\rm const}=2500\,\rm MJ$ per storey was dissipated.
The minimal value for $W_+$ to reach the  data 
point at $t=7.71\sec$ is $W_+=1700\,\rm MJ$ (the blue graph). This corresponds to 
an energy dissipation of $W+W_+=1950\,\rm MJ$ per storey at 
impact height. (The blue graphs in all three diagrams have the same value of $W_+$.)

The two diagrams at the bottom indicate that this value is extremely close to 
arresting the collapse. Indeed the collapse would arrest if this extra force would
continue 10 more meters (or a little more than another  second) as one can see
from the blue graph in the lower right. 
Note that the solution with the constant extra force does 
not arrest if the extra force stays turned on (the magenta graph in the lower right).
The increasing strength of the actual columns is responsible for this effect. 

The diagram to the lower left shows that
the collapse would also arrest if only a slightly bigger value of 1850\,MJ would apply (the yellow graph).
The yellow graph terminates within the errorbars. 
Therefore the distance of the 
crushing front  to the dust front cannot be bigger than the distance from the yellow graph 
to the lower one of the two error bars at time $t=7.71\sec$ (which is less than 5 meters, 
i.\,e.~less than two storeys),
for otherwise the collapse would have terminated.\footnote
{
To be precise at this point: The distance from the dust front to the crushing front could be bigger
than the 5\,m-distance of the yellow solution  to the lower error bar, 
but that would mean that an even higher extra force did occur (over a shorter  interval).
}
A reasonable assumption is that the distance from the crushing front
to the dust front is constant. 
This implies that the solution of the blue graph, which 
does not match the data point at 
$t=9.25\sec$ (in the upper right diagram) is not the solution 
that we are looking for.
But if one increases the extra force a little  ($W_+=1800\,\rm MJ$),
and tuns it off earlier at $318\m$, one obtains the solution given 
in the upper left diagram 
This solution satisfies the empirical requirements. 
This solution also seems to be a better fit because 
from just watching the History Channel clip one might 
guess that the velocity of the dust front is not decelerating 
when it approaches the point $X$. A more refined measurement
could clarify this impression.

Note that the black graph has an energy  dissipation 
that corresponds to an energy dissipation of
$W+W_+=2050\,\rm MJ$ per storey at impact level.

\section{Discussion of Observations}
\subsection{The Magnitude of Energy Dissipation — Revisited}
\label{Subsec-Madonna}
Under the principal assumption that the \red{collapse of the North Tower was gravity-driven—as described by the Crush-Down equation—we} found
that the dissipated energy due to 
column buckling through the first 4.6 seconds
 was on a scale of at most $250\,\rm MJ$ per storey.
In the subsequent three seconds this value increased 
by almost an order of magnitude to over 2000\,MJ. 
After that time period it fell 
back to the initial low value.
(Here we refer to the values $W_+$ relative to the columns at impact height.)

If the maximal possible dissipation of energy per storey is on a scale below 
2000\,MJ  (as it is demanded in \cite{BaLe16}), 
then this  implies that the principal assumption is wrong.

If the maximal possible dissipation of energy can reach the
high values which we determined (as the empirical studies of Korol 
and Sivakumaran indicate \cite{KS14}), then we must urgently 
face the question why this value was not reached 
during the whole of the collapse, i.\,e.~before 4.6 seconds and after 7.7 seconds—either of which
would have terminated the propagation of the collapse.
Understanding the mechanism that enabled this
fluctuation of energy dissipation must have priority in 
a thorough investigation of the collapse. 
In particular, there is no reason whatsoever that one should expect 
that the collapse was inevitable and could not have been arrested 
by the energy dissipation of the buckling columns at any stage 
during the first $8\sec$  of the collapse (and even later).

The numerical values for $\lambda, z_0$ and $\mu_0$ that we used 
are all three smaller than the values used in \cite{BaVe07,BBGL08}.
If we did the same analysis for the higher values therein, our result 
would be even more dramatic in the sense that the additional amount of 
energy dissipation $W_+$ would be bigger. (This statement is obvious for 
$\mu_0$ and also for $z_0$, because less height is compacted.  
To discuss the parameter $\lambda$ note that the  
height of the roof as computed from the measured position of the dust front
decreases with a smaller $\lambda$.)

It would be desirable to have a more refined measurement  
of the downward movement of the crushing front/dust front. 
(We only used two data points.)
As we have determined the camera position for the two clips
from History Channel and CBS, a detailed analysis is possible,
but requires much more effort. (The camera angle is changing, and
the camera is zooming simultaneously.)

\subsection{Conclusion}
This work has presented fundamental empirical data 
of the collapse of the North Tower of the World Trade Center.
These data reveal some highly remarkable phenomena during the collapse: 
\red{Under the assumption of a gravity-driven collapse we have shown that 
the energy dissipation of buckling columns was reduced by an order of magnitude during 
two long time intervalls of the collapse.}
A thorough investigation of the collapse 
is needed to answer the questions that compellingly arise
at this stage.

\begin{appendix}

\section{Determining the Camera Position}
\noindent
\subsection{History Channel}
\label{HistoryApp}
To determine the camera position we compare its perspective with the perspective 
of a known camera position. Other methods are applicable to determine the position, 
however, we  present this method, because it is the most precise one we found.

Consider Figure~\ref{NBCstill}. This is a still from the NBC News
coverage on the 11th of September 2001.
The still is taken at 48\,sec \cite{NBCNews}. The camera is located on the green separation line on West Street. 
The building to the left is the Borough of the Manhattan Community College.
The visible bridge that crosses West Street approximately 100\m southwards
is the Tribeca Bridge (also known as Stuyvesant Bridge).
The big white building is 101 Barclay Street and the tall building behind is WTC\,7.
Note that the camera position is uniquely determined by the position 
of the two street lamps in the picture, which coincidentally happen to be in 
line the north-east corner of 101 Barclay Street, and the north-east corner
of the top floor of the same building, which is also in line with 
the north west corner of WTC\,7.

Figure~\ref{Google} shows a Google Street View screen shot 
of the same location. It is dated January 2013.

Figures~\ref{NYC1}, \ref{NYC2} and \ref{NYC3} show enlarged 
parts of Figure~\ref{NYC}, which is material from an aerial photograph taken in 2006 and 
available on the website of the  City of New York \cite{NYC06}.
The intersection of the blue and the cyan line in these images
is the NBC camera position.
In Figure~\ref{NYC1} the street lamps are visible on the pavement.
They are used for placing the blue and the cyan line.

The length calibration for Figure~\ref{NYC} (the white line of $315.0\m$) 
is set between two
randomly chosen street lamps which have a distance of $315.0\m$.
This length itself is determined by the online tool provided 
by the City of New York (Figure~\ref{NYC0}).

Using figure 
Figure~\ref{FEMA}, which is taken from
 \cite[Chap.\,1, p.\,1-13]{FEMA02},
 we can reconstruct the north-west corner of the North Tower 
 and the north-west corner of WTC\,7 in Figure~\ref{NYC}.
 These are the green and white triangles in the lower part of the images. 
 The thick green line indicates the line following the direction of the 
 north facade of WTC\,7.

Therefore the blue line in Figure~\ref{NYC} measures the 
distance from the NBC camera position to WTC\,7. This length is $d_2=513.1\m$. 
The distance $d_1$  of the north-west corner of WTC\,7 and 
the north east corner of the top floor of 101 Barclay Street 
(the short black line) is 
measured  to be $135.8\m$.
These length measurements have a
precision of  $\pm1\m$.

\begin{center}\begin{figure}
	\includegraphics[scale=0.37, angle=0]{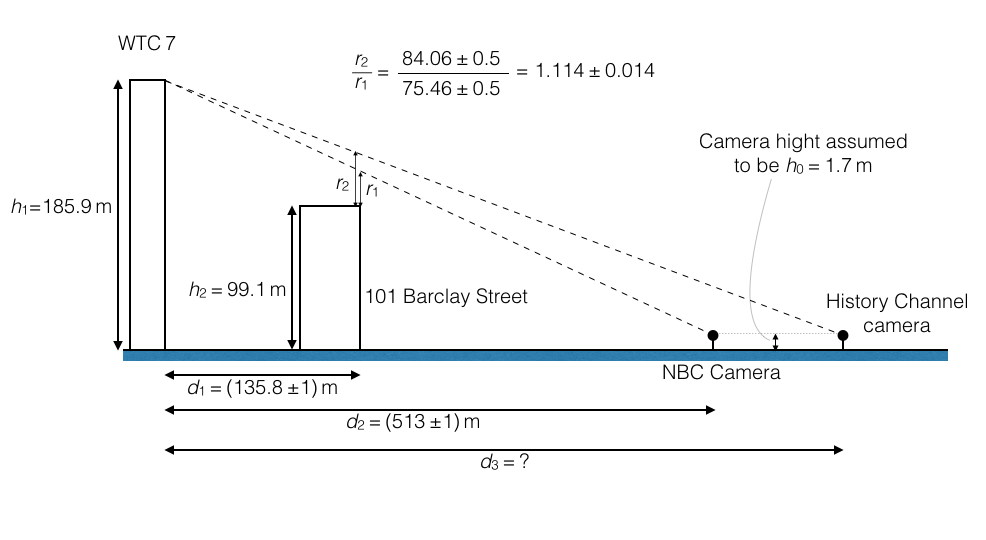}
	\caption{Comparing the two camera positions.}
	\label{CameraPositions}
\end{figure}
\end{center}

Now we are ready to determine the History Channel camera position 
by comparison. Compare Figure~\ref{NBC2} with Figure~\ref{328HistChan}.
Figure~\ref{NBC2} shows a cropped part of the NBC camera image four seconds later (and
less blurred) than Figure~\ref{NBCstill}). Figure~\ref{328HistChan} is a cropped part of
frame 328 of the footage used by History Channel. 
We see that the camera positions are similar but little different.
There is a tiny displacement of the History Channel record to the west and 
a clear displacement to the north, which is recognisable by comparing the indicated vertical measurements. 
The horizontal (black/white) calibration lines are set to 100 reference units. 
Note that the quotient  of the two measured vertical distances is independent of 
the length of the reference unit.
The measurements of the vertical lengths have an error of less than 0.5 reference units.

Because we know the height of WTC\,7 ($h_1=610\,\rm ft=185.9\m$ \cite[p.\,5]{NIST1A}) and the
height of 101 Barclay Street ($h_2=99.06\m$ \cite{Empo}),
we are in the situation illustrated in Figure~\ref{CameraPositions}.
This enables us to determine the distance $d_3$ by the two geometric equations
\begin{eqnarray}
\frac{d_3}{h_1-h_0} =\frac{d_1}{h_1-h_2-r_2}, \quad
\frac{d_2}{h_1-h_0} =\frac{d_1}{h_1-h_2-r_1},
\end{eqnarray}
which gives
\begin{eqnarray}
d_3&=&d_1\left(1-\frac{h_2-h_0}{h_1-h_0}-\frac{r_2}{r_1}\left(1-\frac{h_2-h_0}{h_1-h_0}-\frac{d_1}{d_2} \right) \right)^{-1}
\\\nonumber
\\\nonumber
&=&563\m\pm 9\m.
\end{eqnarray}
With this length the position of the History Channel camera is 
found at the north end of the violett line in Figure~\ref{NYC}.
The distance to the north-west corner of the North Tower (the yellow line in Figure~\ref{NYC}) 
is then 
\begin{eqnarray}
d_0= 694\m\pm 9\m.
\end{eqnarray}

\subsection{CBS}
\label{CBS-Posi}
The position of the  CBS camera can be determined by frame 400 as shown in Figure~\ref{400Unknown} 
up to an ambiguity of $\pm20\m$. It is located on West Street between the two intersections Desbrosses Street and Vestry Street.
For comparison Figure~\ref{GoogleUnknown} shows the same position.  
A distance from the camera to the north-west corner of the North Tower 
of $(1202\pm20)\m$ is measured in Figure~\ref{NYC-Unknown}.

\section{Synchronising the Video Clips}
\label{AppTime}
\subsection{History Channel}
\begin{figure}[h]
	\includegraphics[scale=0.2]{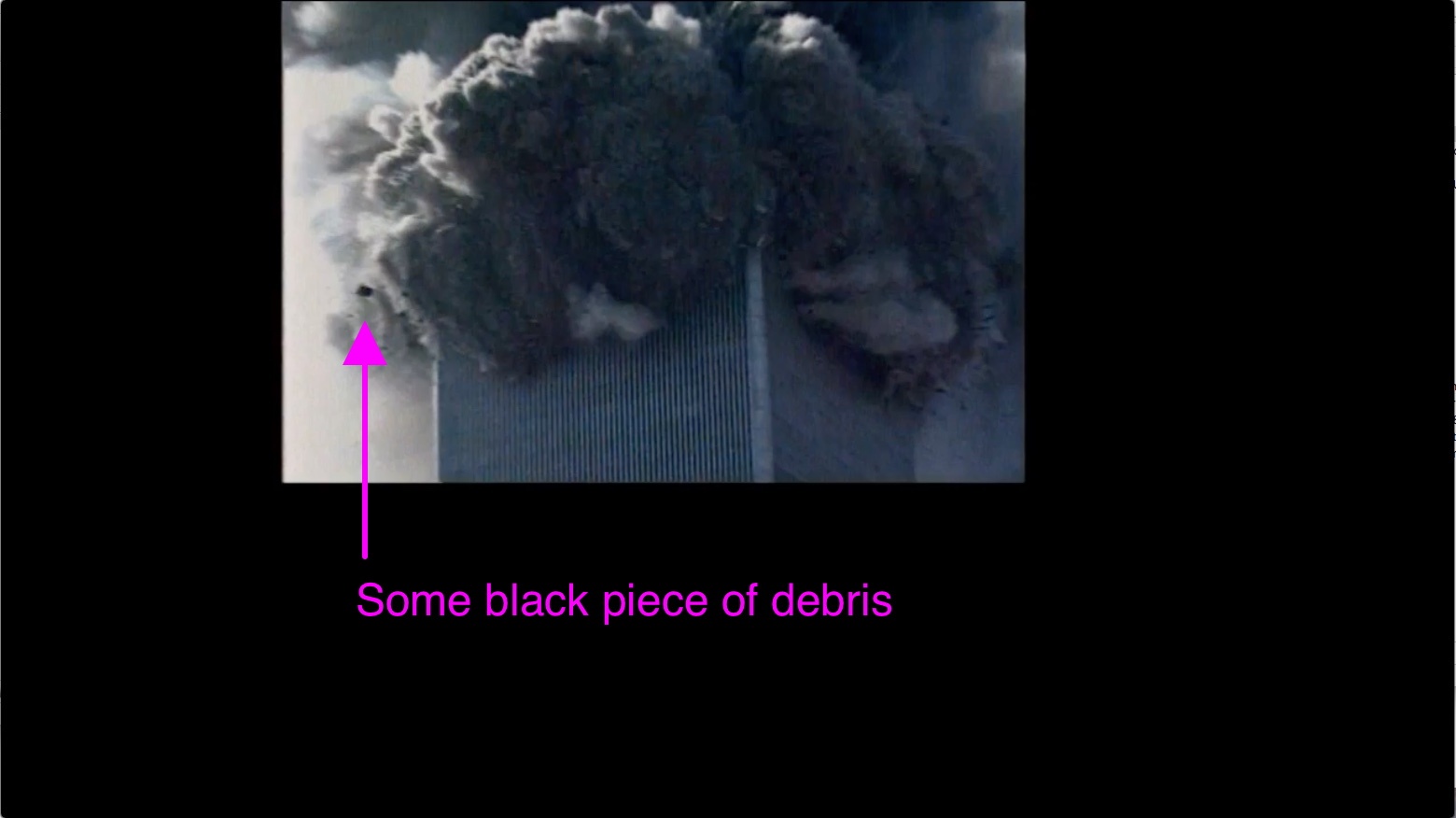}
	\caption{\red{The first frame of the History Channel clip.}}
	\label{0HistChan}
\end{figure}

There are some rough methods to pre-adjust the 
Sauret clips and the History Channel clip up to a third of a second. 
E.\,g. there is a clearly recognisable black part of debris falling left (east) 
to the tower, which is on the first frame of the History Channel video \red{(Fig.~\ref{0HistChan})}.
In the Sauret Video this very same piece of debris is visible 
from approximately  $t=3.2\sec$ to $t=3.7\sec$ in the lower left \red{(Fig.~\ref{1030Sauret})}.

Once a rough calibration is done we are looking for an event that 
can be used to a synchronisation  up to one frame. There are plenty of those:
In the dust there are numerous and well localisable `blinks' appearing. 
Some of them are only visible for one frame.

Not all of the blinks are visible in all camera perspectives, but
we use such one for synchronisation that is visible in at least three 
records. The third record we use is taken from a CNN documentary \cite{CNN}
and gives an intermediate perspective between Sauret's camera and 
the camera on West Street.
Figureres~\ref{289SauretBlink} to \ref{111HistoryBlink} 
show the disappearance of the same blink in the three perspectives from one frame 
to the next. This is 1089 to 1090 in Sauret's video, 110 to 111 in the History Channel 
video and 127 to 128 in the CNN video. 
We therefore identify the timeline 
of these videos at this step. 
For controlling reasons we have verified this synchronisation with other blinks and 
found confirmation up to one frame.
This is as good as it can possibly be. 
Note the History Channel clip has the double frame rate 
of Sauret's video. 

Consequently, the time of frame $262$ of the History Channel clip 
is given by the time of frame $1165\pm1$ in Sauret's video. 
This time is $t=7.71\sec\pm 0.03\sec$ after collapse initiation (at frame 934).

\subsection{CBS}
\label{CBS-Sync}
The blink that has been used to synchronise the Sauret video and the History Channel clip
is not clearly visible in the CBS clip. But the appearance of another blink (Blink 2) in Figures~\ref{109HistoryBlink} to \ref{636CBS}
can be used to synchronise the History Channel clip and the CBS clip:
The step from frame 636 to 637 (CBS) corresponds to 109 to 110 (History Channel),
which happens at time $t=5.17$.

The frame rate of the CBS clip is 25 frames per second.
Therefore fame 739 of the CBS clip is $4.08\sec$ after frame
637. This is $t=9.25\sec$ after collapse initiation.

\section{Computing   Numerical Solutions with Maxima}
\label{Numerics}
\noindent
The following is the  source code which we have used to compute the 
solutions of the Crush-Down Equation with Maxima \cite{Maxi}. 
The variable \texttt{z\_1} corresponds to $29\,h$. 
\texttt{v\_0:0} sets the initial velocity to zero.
Note for the computation that
the mass density $\mu_0$, the parameter $\beta$ and the energy absorption capacity of the columns $W$ miss a factor $10^6$ 
in the source code. 
However, 
this factor cancels out in the coefficients $\phi$ and $\psi$, so the solution is not effected 
by this simplification.

{\tiny
\texttt{
\\/* [wxMaxima: input   start] */\\
/* [Define the constants] */\\
mu\_0:0.6; g:9.8; H:417; h:3.8; z\_0:46; z\_1:110;v\_0:0;
\\\\
/* [We compute 4 soltuions so we give the following parameters fourfold] */\\\\
lambda\_1:0.15;kappa\_1:0.25;\\ 
lambda\_2:0.15;kappa\_2:0.25; \\
lambda\_3:0.15;kappa\_3:0.25; \\
lambda\_4:0.15;kappa\_4:0.25; \\
\\
alpha\_1:0.1*kappa\_1*(1-lambda\_1)\^{}2; beta\_1:0.05;\\
alpha\_2:0.1*kappa\_2*(1-lambda\_2)\^{}2; beta\_2:0.05;\\
alpha\_3:0.1*kappa\_3*(1-lambda\_3)\^{}2; beta\_3:0.05;\\
alpha\_4:0.1*kappa\_4*(1-lambda\_4)\^{}2; beta\_4:0.05;\\
\\
W\_1:250;
W\_2:250;
W\_3:250;
W\_4:250;
\\\\
/* [The measured data] */\\\\
t\_1:1.64; a\_1:408; error\_1:2;\\
t\_2:3.20; a\_2:383; error\_2:2;\\
t\_3:4.57; a\_3:349; error\_3:4;\\
t\_4:7.71; a\_4:248; error\_4:2;\\
t\_5:9.25; a\_5:216; error\_5:1.5;\\
\\
/* [The Heaviside step function] */\\\\
theta(z):=if z<0 then 0 else 1;
\\\\
/* [The damage function] */\\\\
chi(z):=(0.5+0.4*theta(z-z\_0-h)+0.1*theta(z-z\_0-4*h));
\\\\
/* [The mass density and the mass function] */\\\\
mu(z):= mu\_0*(1+theta(z-z\_1)*(0.43*(z-z\_1)/(H-z\_1)));
\\\\
m\_1(z):= mu\_0*z\_0+ (1-kappa\_1)*mu\_0*(z-z\_0+ theta(z-z\_1)*0.215*(z-z\_1)\^{}2/(H-z\_1));\\
m\_2(z):= mu\_0*z\_0+ (1-kappa\_2)*mu\_0*(z-z\_0+ theta(z-z\_1)*0.215*(z-z\_1)\^{}2/(H-z\_1));\\
m\_3(z):= mu\_0*z\_0+ (1-kappa\_3)*mu\_0*(z-z\_0+ theta(z-z\_1)*0.215*(z-z\_1)\^{}2/(H-z\_1));\\
m\_4(z):= mu\_0*z\_0+ (1-kappa\_4)*mu\_0*(z-z\_0+ theta(z-z\_1)*0.215*(z-z\_1)\^{}2/(H-z\_1));\\
\\
/* [The amount of extra energy dissipation] */\\\\
W\_extra\_1:1000;
W\_extra\_2:1500;
W\_extra\_3:2000;
\\\\
/* [Turning the forces On and Off] */\\\\
on:a\_3+error\_3+10; 
off:a\_4-error\_4+z\_0+lambda\_1*(H-z\_0-(a\_4-error\_4) ); 
\\\\
On\_1(z):=theta(z-(z\_0+(H-on)/(1-lambda\_1))); Off\_1(z):=-theta(z-(z\_0+(H-off)/(1-lambda\_1)));\\
On\_2(z):=theta(z-(z\_0+(H-on)/(1-lambda\_2))); Off\_2(z):=-theta(z-(z\_0+(H-off)/(1-lambda\_2)));\\
On\_3(z):=theta(z-(z\_0+(H-on)/(1-lambda\_3))); Off\_3(z):=-theta(z-(z\_0+(H-off)/(1-lambda\_3)));\\
\\
/* [The extra forces] */\\\\
Extra\_1(z):= W\_extra\_1/h * (On\_1(z)+Off\_1(z));\\
Extra\_2(z):= W\_extra\_2/h * (On\_2(z)+Off\_2(z));\\
Extra\_3(z):= W\_extra\_3/h * (On\_3(z)+Off\_3(z));\\
\\
/* [The total upward column force] */\\\\
F\_1(z):= (Extra\_1(z)+W\_1/h)*(1+theta(z-z\_1)*(6*(z-z\_1)/(H-z\_1)));\\
F\_2(z):= (Extra\_2(z)+W\_2/h)*(1+theta(z-z\_1)*(6*(z-z\_1)/(H-z\_1)));\\
F\_3(z):= (Extra\_3(z)+W\_3/h)*(1+theta(z-z\_1)*(6*(z-z\_1)/(H-z\_1)));\\
F\_4(z):=          		  (W\_4/h)*(1+theta(z-z\_1)*(6*(z-z\_1)/(H-z\_1)));\\
\\
/* [The coefficients of the Crush-Down Equation] */\\\\
phi\_1(z):=g/(1-lambda\_1)-chi(z)*F\_1(z)/((1-lambda\_1)*m\_1(z));\\
phi\_2(z):=g/(1-lambda\_2)-chi(z)*F\_2(z)/((1-lambda\_2)*m\_2(z));\\
phi\_3(z):=g/(1-lambda\_3)-chi(z)*F\_3(z)/((1-lambda\_3)*m\_3(z));\\
phi\_4(z):=g/(1-lambda\_4)-chi(z)*F\_4(z)/((1-lambda\_3)*m\_4(z));\\
\\
psi\_1(z):=(1-kappa\_1)*mu(z)/m\_1(z) + (alpha\_1*mu(z)+beta\_1)/((1-lambda\_1)*m\_1(z));\\
psi\_2(z):=(1-kappa\_2)*mu(z)/m\_2(z) + (alpha\_2*mu(z)+beta\_2)/((1-lambda\_2)*m\_2(z));\\
psi\_3(z):=(1-kappa\_3)*mu(z)/m\_3(z) + (alpha\_3*mu(z)+beta\_3)/((1-lambda\_3)*m\_3(z));\\
psi\_4(z):=(1-kappa\_4)*mu(z)/m\_4(z) + (alpha\_4*mu(z)+beta\_4)/((1-lambda\_4)*m\_4(z));\\
\\
/* [Compute the solutions with Runge-Kutta] */\\
/* [Using the Heaviside function the solutions arrest at negative velocities] */\\\\
time:12;stepwidth:0.01;\\
solution\_1:rk ( [u* theta(u), phi\_1(z)-u\^{}2*psi\_1(z)], [z, u], [z\_0,v\_0], [t,0,time,stepwidth])\$\\
solution\_2:rk ( [u* theta(u), phi\_2(z)-u\^{}2*psi\_2(z)], [z, u], [z\_0,v\_0], [t,0,time,stepwidth])\$\\
solution\_3:rk ( [u* theta(u), phi\_3(z)-u\^{}2*psi\_3(z)], [z, u], [z\_0,v\_0], [t,0,time,stepwidth])\$\\
solution\_4:rk ( [u* theta(u), phi\_4(z)-u\^{}2*psi\_4(z)], [z, u], [z\_0,v\_0], [t,0,time,stepwidth])\$\\
\\
/* [Turn the solutions into the height of the roof] */\\\\
height\_1:makelist([solution\_1[i][1],H-(1-lambda\_1)*(solution\_1[i][2]-z\_0)],i,1,length(solution\_1))\$\\
height\_2:makelist([solution\_2[i][1],H-(1-lambda\_2)*(solution\_2[i][2]-z\_0)],i,1,length(solution\_2))\$\\
height\_3:makelist([solution\_3[i][1],H-(1-lambda\_3)*(solution\_3[i][2]-z\_0)],i,1,length(solution\_3))\$\\
height\_4:makelist([solution\_4[i][1],H-(1-lambda\_4)*(solution\_4[i][2]-z\_0)],i,1,length(solution\_4))\$\\
\\
/* [Plot the solutions and the empirical data] */\\\\
wxplot2d( [ \\{}[discrete,height\_1], [discrete,height\_2],
           	[discrete,height\_3],  [discrete,height\_4],\\{}\\{}
             {}[parametric, t\_1, t, [t, 0, a\_1]],\\{}[parametric,t, a\_1+error\_1, [t, 0,t\_1]],\\{}[parametric,t, a\_1-error\_2, [t, 0,t\_1]],\\\\{}
          {}  [parametric, t\_2, t, [t, 0, a\_2]],\\{}[parametric,t, a\_2+error\_2, [t, 0,t\_2]],\\{}[parametric,t, a\_2-error\_2, [t, 0,t\_2]],\\\\{}
        {}    [parametric, t\_3, t, [t, 0, a\_3]],\\{}[parametric,t, a\_3+error\_3, [t, 0,t\_3]],\\{}[parametric,t, a\_3-error\_3, [t, 0,t\_3]],\\\\{}
            \\
         {}   [parametric, t\_4, t, [t, 0, a\_4+z\_0+lambda\_1*(H-z\_0-a\_4)]],\\
    {}        [parametric,t, a\_4-error\_4+z\_0+lambda\_1*(H-z\_0-(a\_4-error\_4)),[t, 0,t\_4]],\\
      {}      [parametric,t, a\_4+error\_4+z\_0+lambda\_1*(H-z\_0-(a\_4+error\_4)),[t, 0,t\_4]],\\
            \\
      {}      [parametric, t\_5, t, [t, 0, a\_5+z\_0+lambda\_1*(H-z\_0-a\_5)]],\\
      {}      [parametric,t, a\_5-error\_5+z\_0+lambda\_1*(H-z\_0-(a\_5-error\_5)),[t, 0,t\_5]],\\
        {}    [parametric,t, a\_5+error\_5+z\_0+lambda\_1*(H-z\_0-(a\_5+error\_5)),[t, 0,t\_5]]\\
            ],\\{}
            [x,0,time], [y,260,420],\\
      {}      [style,[lines,1,orange],
                               [lines,1,cyan],
                               [lines,1,black],
                               [lines,1,red],
                               [dots,black],
                               [lines,1,black],
                               [lines,1,black],
                               [dots,black],
                               [lines,1,black],
                               [lines,1,black],
                               [dots,black],
                               [lines,1,black],
                               [lines,1,black],
                               [dots,black],
                               [lines,1,black],
                               [dots,black],
                               [dots,black],
                               [lines,1,black],
                               [dots,black],
                               [lines,1,black]
    ],\\
   {}     [ylabel,"Height of tower top / m "], [xlabel,"Time / sec"], \\{}
            [title,concat("Extra Force On: ", string(on)," m, Off: ", string(off)," m" ) ],\\{}
            [legend,
                    concat("W\_+=",string(W\_extra\_1)," MJ"),
                    concat("W\_+=",string(W\_extra\_2)," MJ"),
                    concat("W\_+=",string(W\_extra\_3)," MJ"),
                    concat("W=",string(W\_4)," MJ"),\\{}
                    "","","","","","","","","","","","","","","",""]\\)\$\\\\
/* [wxMaxima: input   end] */
}
}

\newpage
\begin{figure}
	\includegraphics[scale=0.3, angle=90]{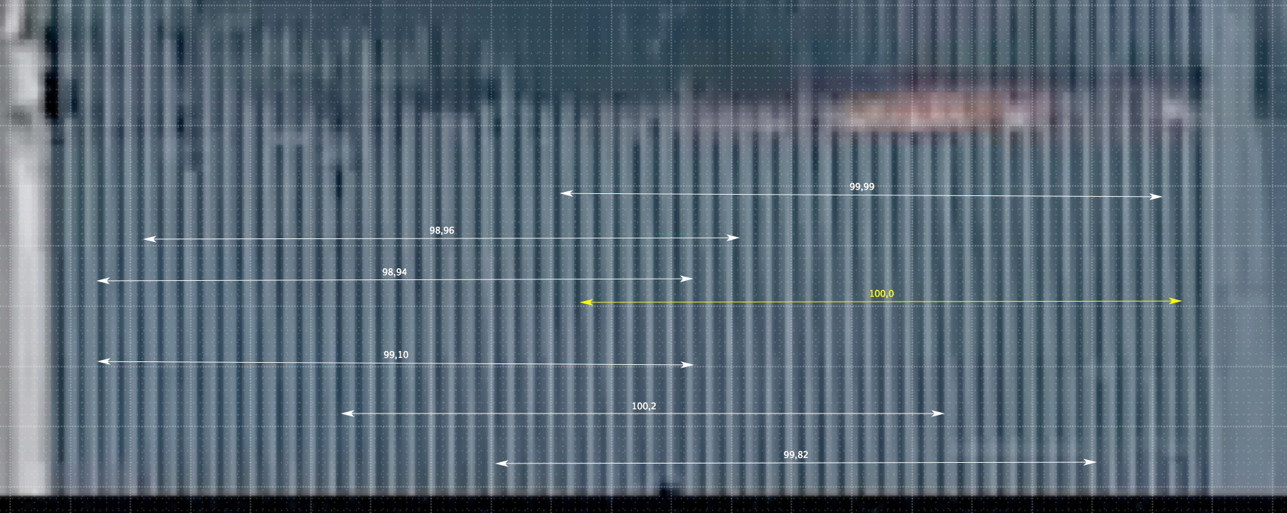}
	\caption{Horizontal calibration measurement at frame 800, $t=-4.47\sec$.}
	\label{800SauretZoom}
\end{figure}

\begin{figure}
	\includegraphics[scale=0.47]{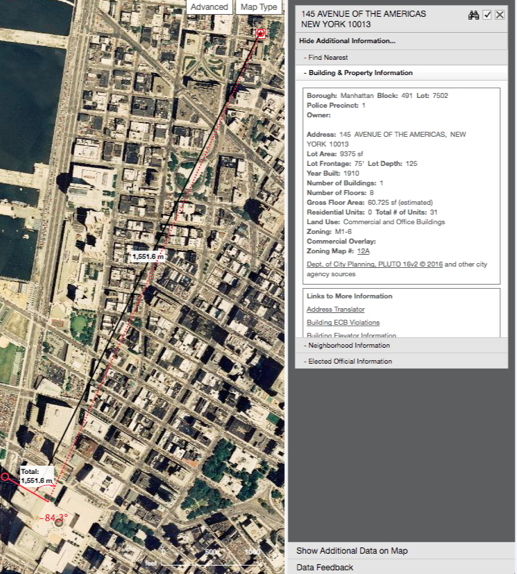}
	\caption{Distance from 145 Avenue of the Americas to the WTC~complex based on an aerial photograph from 1996. 
	The distance measurement is done with the provided online tool of the  City of New York \cite{NYC96}.
	Note for the angle measurement that the optical center of the camera points little eastwards  to the building.}
	\label{145Avenue}
	\includegraphics[scale=0.195]{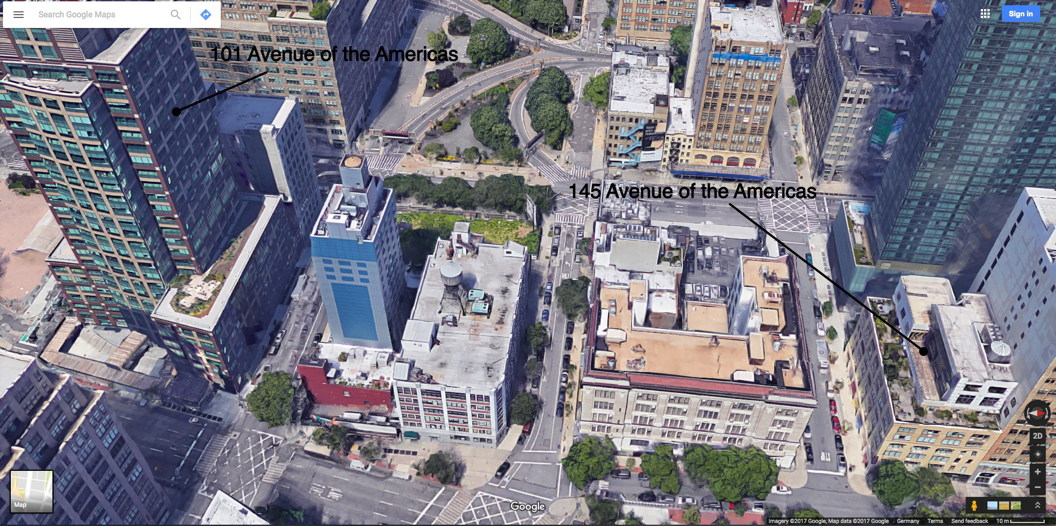}
	\caption{Screenshot from Google Street View showing 101 Avenue of the Americas and 145 Avenue of the Americas. 
	\cite{GSV17}.}
	\label{Google145Ave}
\end{figure}

\begin{figure}
	\includegraphics[scale=0.42,angle=0]{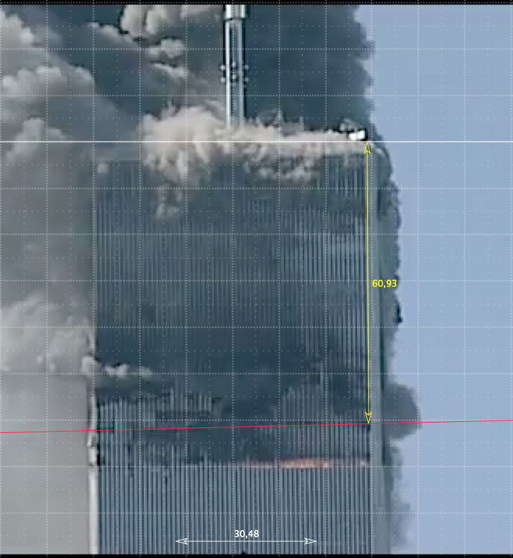}
	\caption{Measuring a known vertical distance, $t=-4.47\sec$.}
	\label{800SauretVertCalibration}

	\includegraphics[scale=0.3,angle=0]{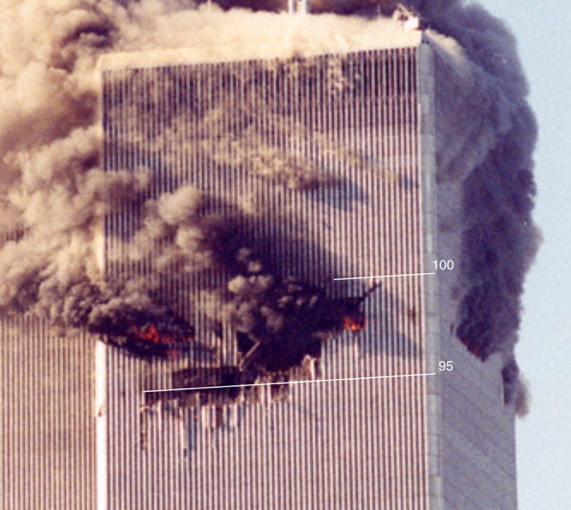}
	\caption{North side of the North Tower, \cite[p.\,35]{NIST1-5A}. The white lines and the floor number are added 
	on the basis of Figure~\ref{p22-1}.}
	\label{p35-1-5A}
\end{figure}

\begin{figure}
	\includegraphics[scale=0.34,angle=0]{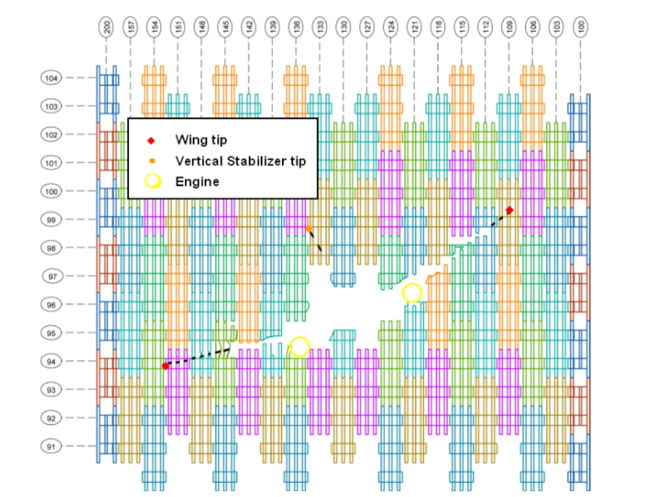}
	\caption{Schematic illustration of the aircraft impact zone \cite[p.\,22]{NIST1}.}
	\label{p22-1}

	\includegraphics[scale=0.34,angle=0]{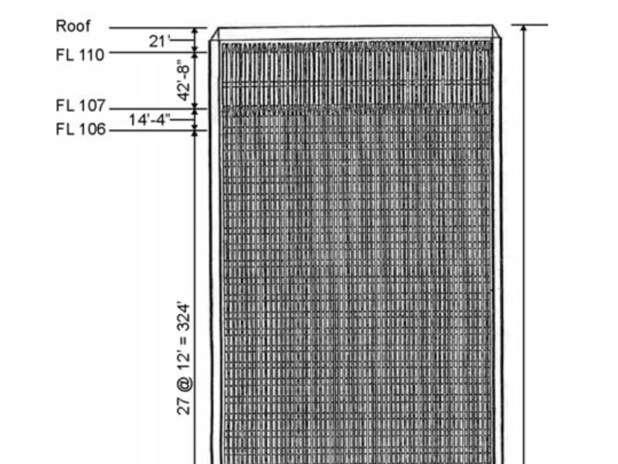}
	\caption{Structural drawing of the North Tower \cite[p.\,18]{NIST1-1}.}
	\label{p18-1-1}

\end{figure}

\begin{figure}
	\includegraphics[scale=0.4,angle=0]{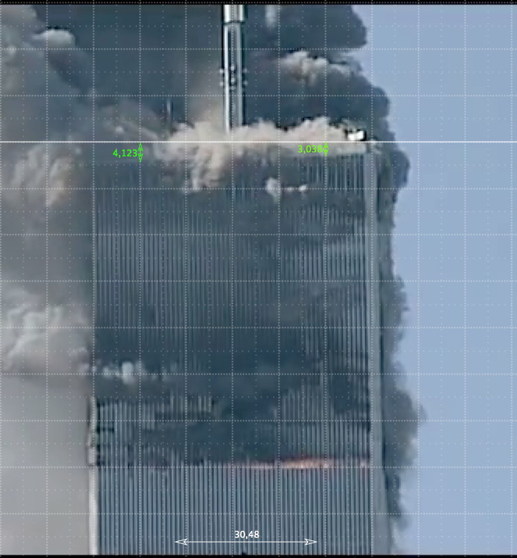}
	\caption{Measuring the  position of the bottom of the roof at frame 934, $t=0\sec$.}
	\label{934Sauret}

	\includegraphics[scale=0.4,angle=0]{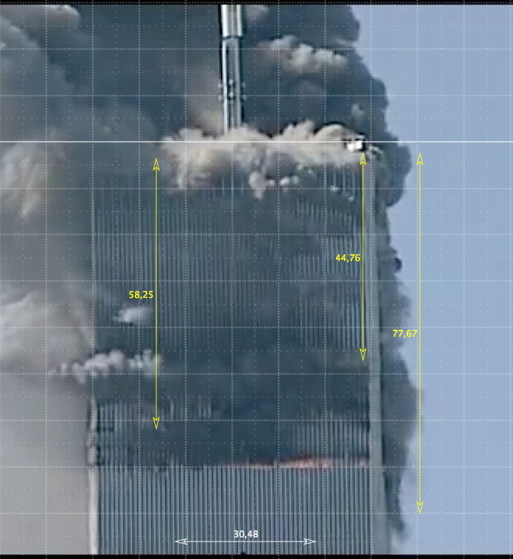}
	\caption{Measuring the height of the initially falling block at frame 957, $t=0.77\sec$.}
	\label{957Sauret}
\end{figure}

\begin{figure}

	\includegraphics[scale=0.4, angle=0]{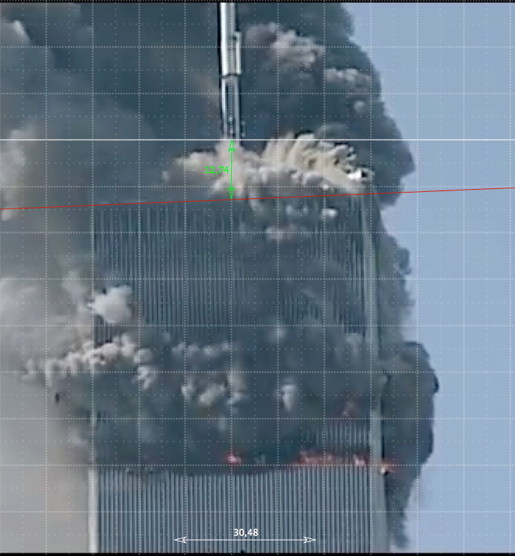}
	\caption{Measuring the position of the bottom of the roof at frame 983, $t=1.64\sec$.}
	\label{983Sauret}

	\includegraphics[scale=0.4,angle=0]{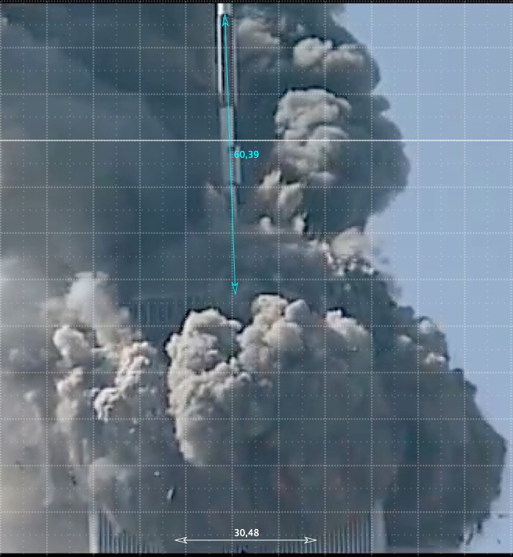}
	\caption{Measuring the lower part of the antenna at frame 1024, $t=3.00\sec$.}
	\label{1024Sauret}
\end{figure}
\begin{figure}

	\includegraphics[scale=0.4, angle=0]{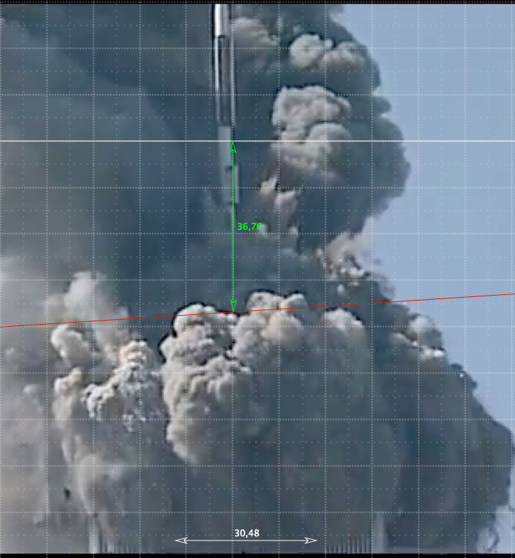}
	\caption{Measuring the position of the bottom of the roof at frame 1030, $t=3.20\sec$.}
	\label{1030Sauret}

	\includegraphics[scale=0.4]{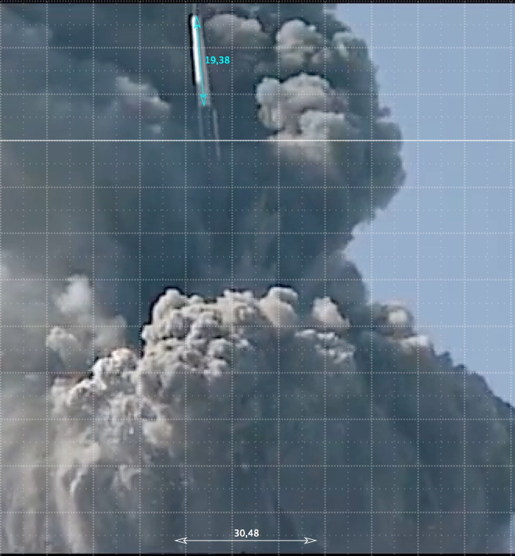}
	\caption{Measuring the white part of the antenna at frame 1050, $t=3.87\sec$.}
	\label{1050Sauret}
	
\end{figure}
\begin{figure}
	\includegraphics[scale=0.4]{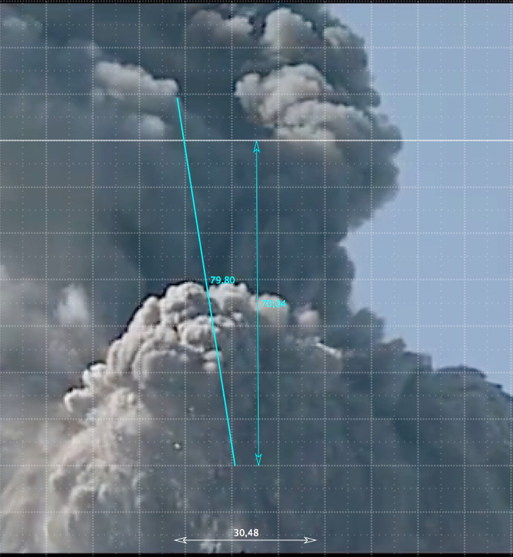}
	\caption{Reconstructing the position of the bottom of the roof at frame 1071, $t=4.57\sec$.}
	\label{1071Sauret}
\end{figure}

\begin{center}\begin{figure}
	\includegraphics[scale=0.36]{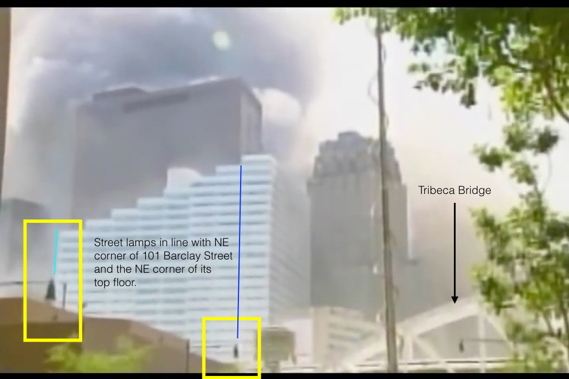}
	\caption{The NBC News camera perspective at 48\,sec \cite{NBCNews}.}
	\label{NBCstill}
\end{figure}
\end{center}

\begin{center}

\begin{figure}
	\includegraphics[scale=0.363, angle=0]{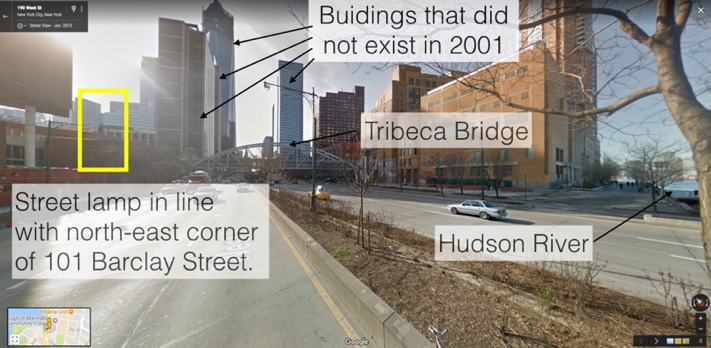}
	\caption{Screen shot from Google Street View, showing West Street in January 2013, \cite{GSV13}.}
	\label{Google}
\end{figure}
\end{center}

\begin{center}\begin{figure}
	\includegraphics[scale=0.7]{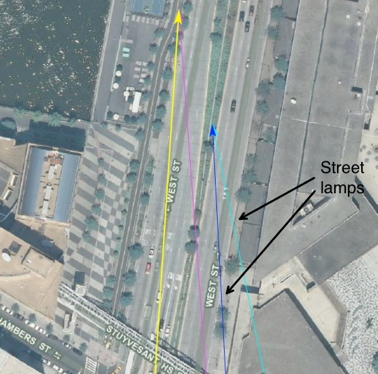}
	\caption{Zoom into the top part of Figure~\ref{NYC}.}
	\label{NYC1}
\end{figure}
\end{center}

\begin{center}\begin{figure}
	\includegraphics[scale=0.7]{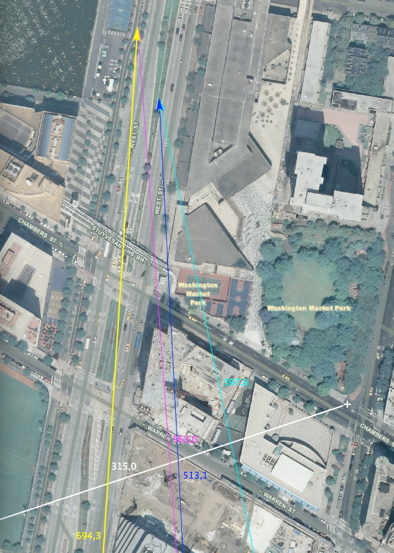}
	\caption{Zoom into the upper part of Figure~\ref{NYC}.}
	\label{NYC2}
\end{figure}
\end{center}

\begin{center}\begin{figure}
	\includegraphics[scale=0.78, angle=0]{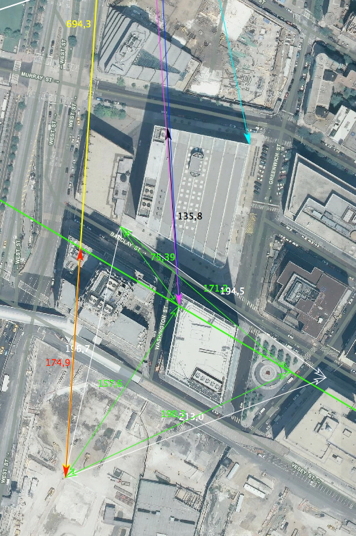}
	\caption{Zoom into the lower part of Figure~\ref{NYC}.}
	\label{NYC3}
\end{figure}
\end{center}

\begin{center}\begin{figure}
	\includegraphics[scale=0.67, angle=0]{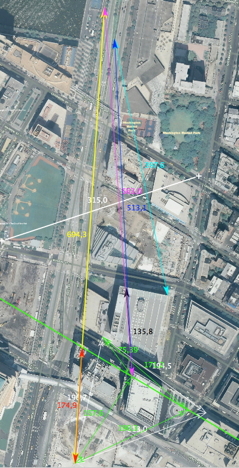}
	\caption{Aerial photograph of NYC, dated 2006, \cite{NYC06}. The green and white measurements 
	at the bottom indicate the reconstruction of the position of the NW corner of the North Tower and of the 
	north face of WTC\,7 (cp.~Figure~\ref{FEMA}). After having reconstructed the 
	the position of the NW corner we noticed the clearly recognisable ground formation that forms a right angle 
	where we determined the corner (cp.~Figure~\ref{NYC3}).  This might be remains of the actual foot print 
	of the tower, which indicates 
	that the reconstruction is done properly.}
	\label{NYC}
\end{figure}
\end{center}

\begin{center}\begin{figure}
	\includegraphics[scale=0.47, angle=90]{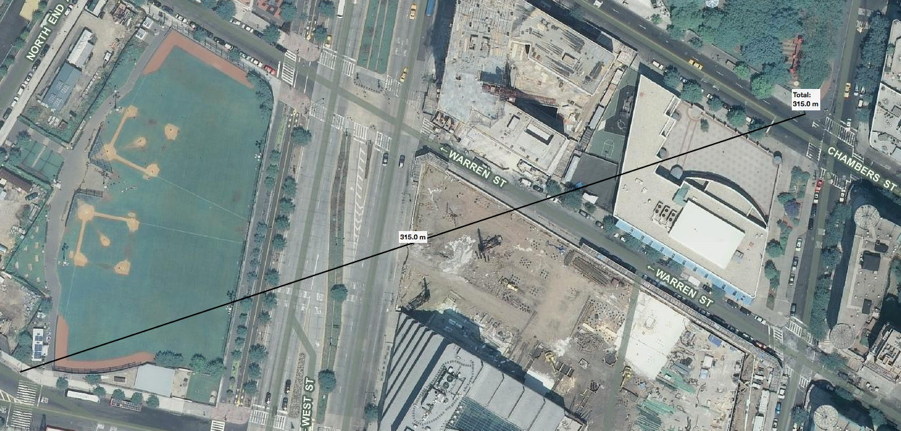}
	\caption{Calibration measurements for Figure~\ref{NYC} using the online
	measurement tool of \cite{NYC06}.}
	\label{NYC0}
\end{figure}
\end{center}

\begin{center}
\begin{figure}
	\includegraphics[scale=0.45, angle=0]{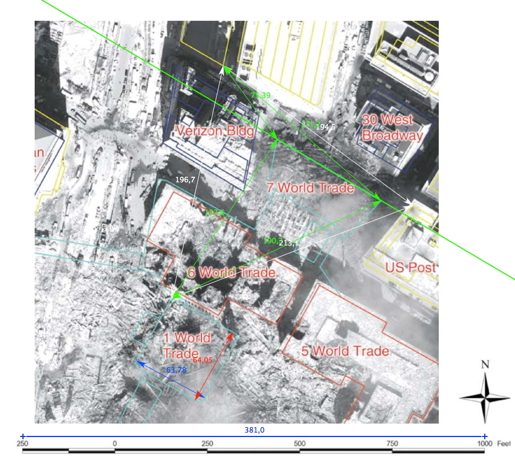}
	\caption{Aerial photograph, taken from \cite[Ch.\,1, p.\,1-13]{FEMA02}. Using the displayed  
	measurements one can reconstruct the position of the north west corner of the North Tower and the 
	north side of WTC\,7 in Figure~\ref{NYC}
	by just transporting all triangles.}
	\label{FEMA}
\end{figure}
\end{center}

\begin{center}
\begin{figure}
	\includegraphics[scale=0.47, angle=0]{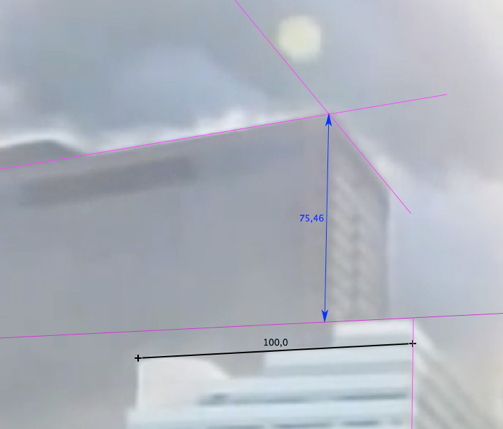}
	\caption{NBC News camera perspective (cropped) at 52\,sec.}
	\label{NBC2}
\end{figure}
\end{center}

\begin{center}
\begin{figure}
	\includegraphics[scale=0.34, angle=0]{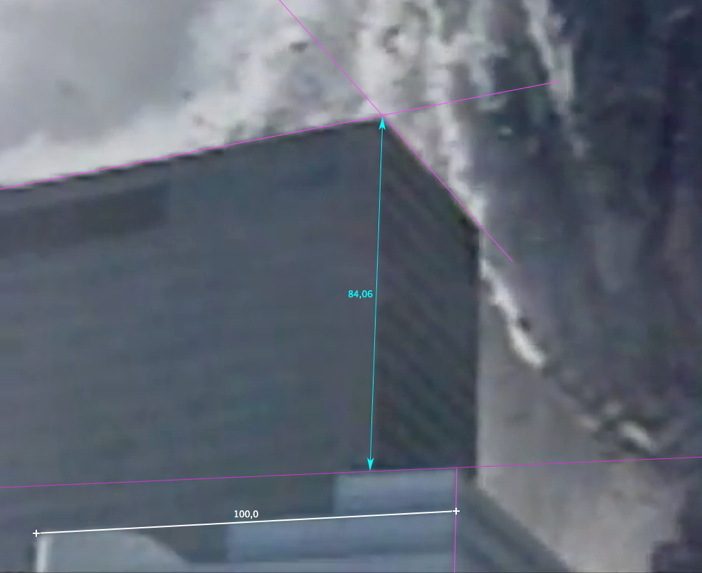}
	\caption{History Channel camera perspective (cropped), frame 328.}
	\label{328HistChan}
\end{figure}
\end{center}

\begin{figure}
	\includegraphics[scale=0.6, angle=0]{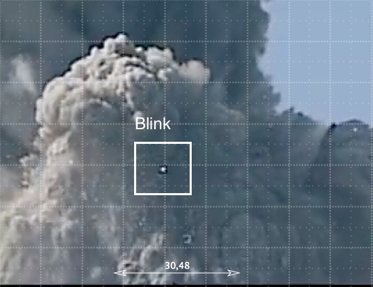}
	\caption{Zoom into frame 1089 of Sauret's video.}
	\label{289SauretBlink}
\end{figure}

\begin{figure}
	\includegraphics[scale=0.6, angle=0]{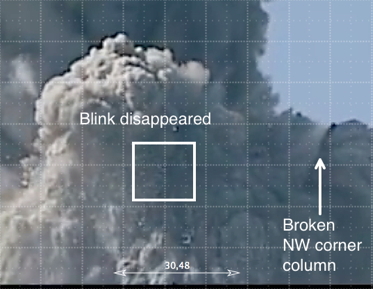}
	\caption{Zoom into frame 1090 of Sauret's video.}
	\label{290SauretBlink}
\end{figure}

\begin{figure}
	\includegraphics[scale=0.44, angle=0]{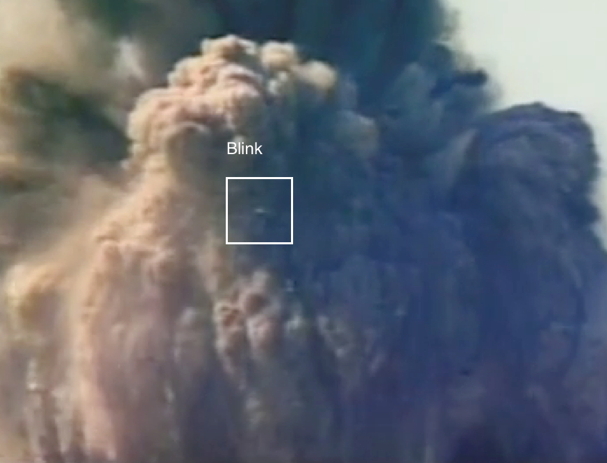}
	\caption{For comparison: Zoom into frame 127 of the CNN video.}
	\label{127CNN}
\end{figure}

\begin{figure}
	\includegraphics[scale=0.4, angle=0]{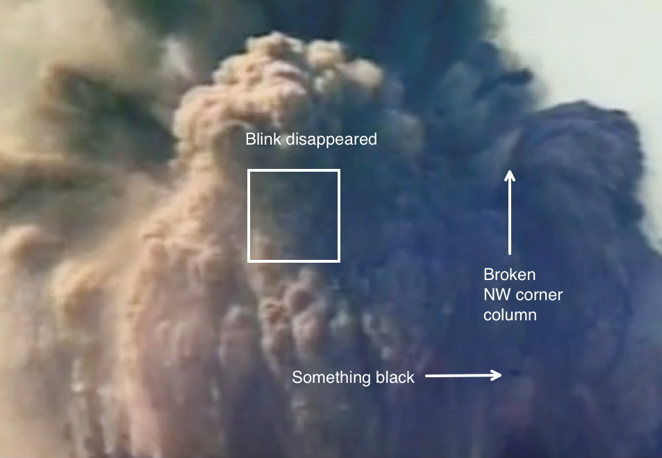}
	\caption{For comparison: Zoom into frame 128 of the CNN video.}
	\label{128CNN}
\end{figure}

\begin{figure}
	\includegraphics[scale=0.33, angle=0]{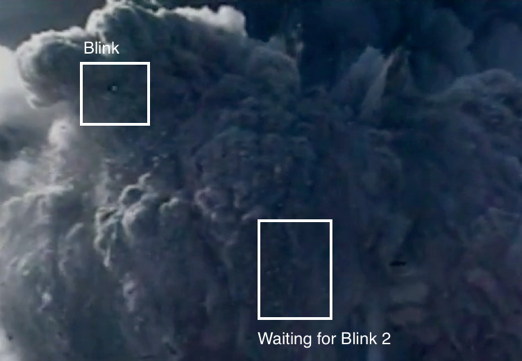}
	\caption{Zoom into frame 109 of the History Channel video.}
	\label{109HistoryBlink}
\end{figure}

\begin{figure}
	\includegraphics[scale=0.26, angle=0]{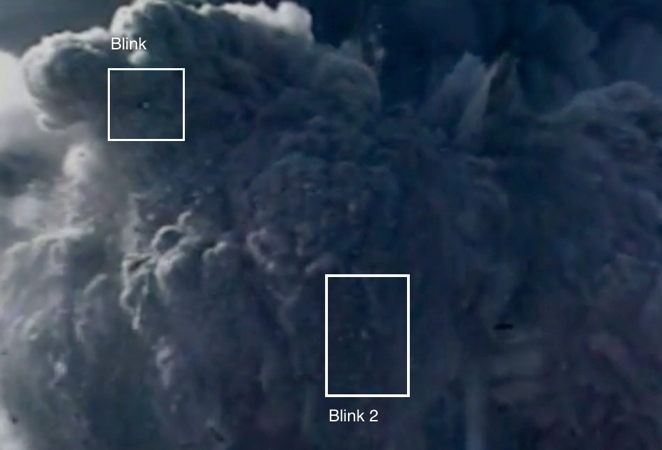}
	\caption{Zoom into frame 110 of the History Channel video.}
	\label{110HistoryBlink}
\end{figure}

\begin{figure}
	\includegraphics[scale=0.26, angle=0]{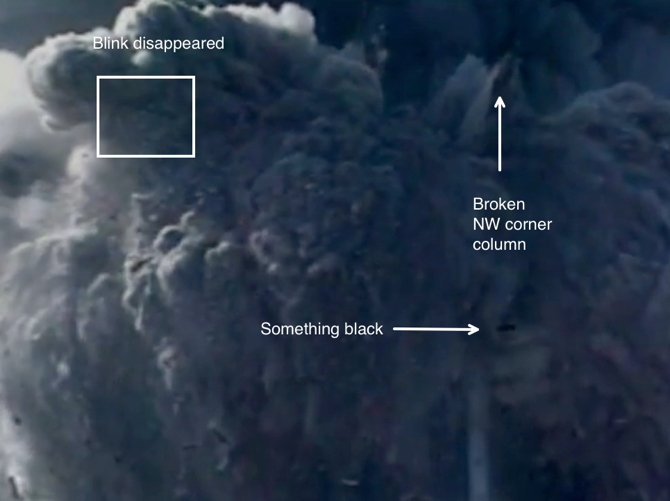}
	\caption{Zoom into frame 111 of the History Channel video.}
	\label{111HistoryBlink}
\end{figure}

\begin{figure}
	\includegraphics[scale=0.4, angle=0]{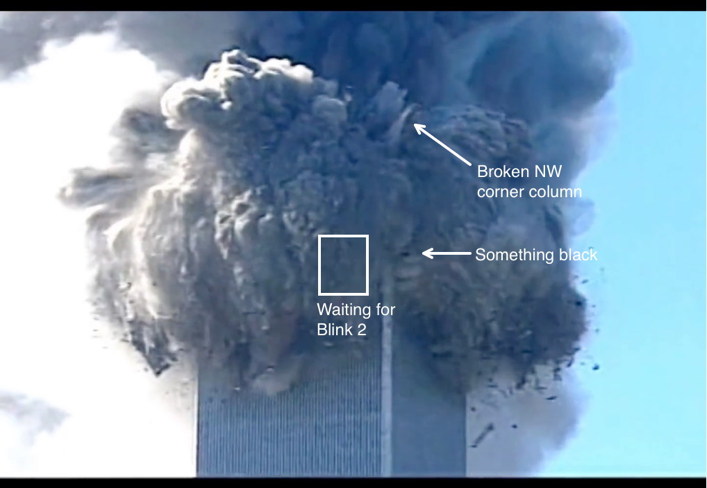}
	\caption{Frame 636 of the CBS clip.}
	\label{636CBS}
\end{figure}

\begin{figure}
	\includegraphics[scale=0.4, angle=0]{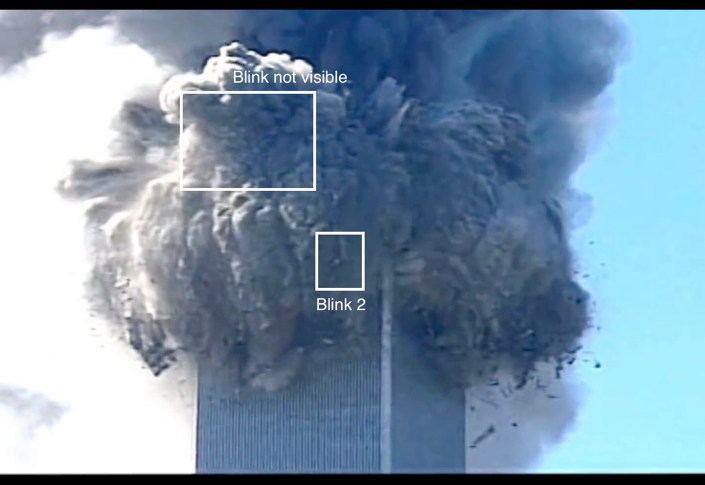}
	\caption{Frame 637 of the CBS clip.}
	\label{637CBS}
\end{figure}

\begin{figure}
	\includegraphics[scale=0.334, angle=0]{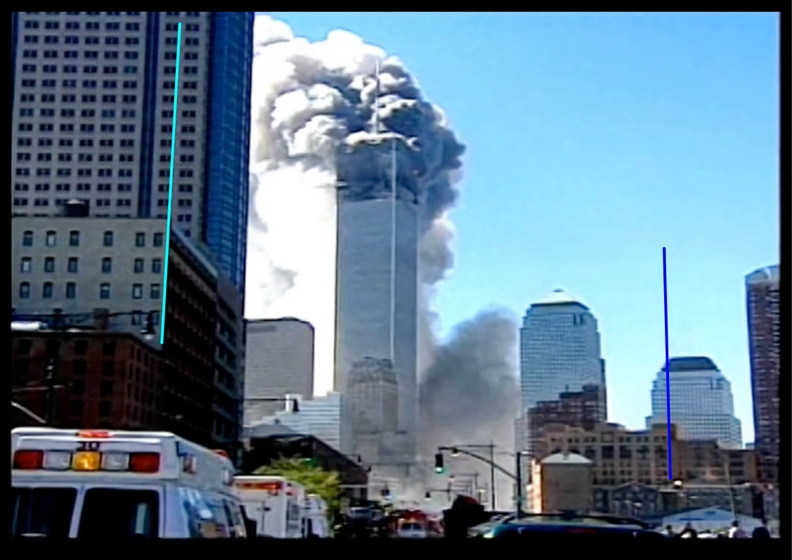}
	\caption{Frame 400 of the CBS clip \cite{CBS01}.}
	\label{400Unknown}
\end{figure}

\begin{figure}

	\includegraphics[scale=0.22, angle=0]{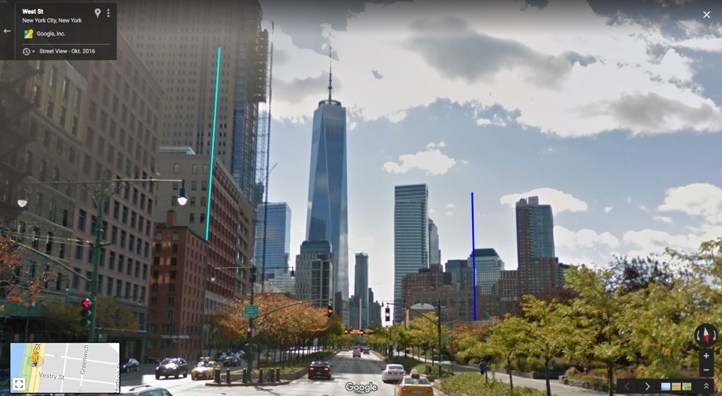}
	\caption{Screen shot from Google Street View, showing West Street in October 2016, \cite{GSV16}.}
	\label{GoogleUnknown}
\end{figure}

\begin{figure}
	\includegraphics[scale=0.7]{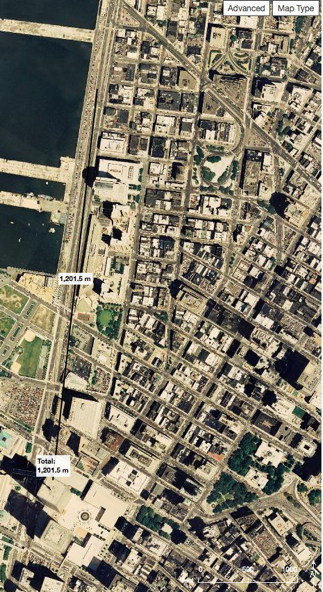}
	\caption{Distance from the CBS camera position to the WTC~complex based on an aerial photograph from 1996. 
	The measurement is done with the provided online tool of the  City of New York \cite{NYC96}.}
	\label{NYC-Unknown}
\end{figure}

\begin{figure}
	\includegraphics[scale=0.45]{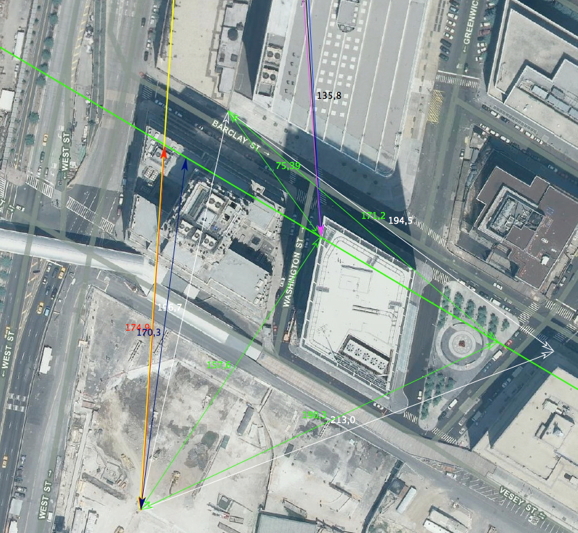}
	\caption{The dark blue line indicates the distance 
	from the north-west corner of the tower to the intersection of the green line in the direction of the CBS camera.}
	\label{739-CBS-WTC7}
\end{figure}

\end{appendix}

\end{document}